\documentclass[11pt]{article}

\usepackage[margin=1in]{geometry}

\usepackage[utf8]{inputenc} 
\usepackage[T1]{fontenc}    
\usepackage{url}            
\usepackage{booktabs}       
\usepackage{amsfonts}       
\usepackage{nicefrac}       
\usepackage{microtype}      
\usepackage{xcolor}         
\usepackage{amsthm}
\usepackage{hyperref}       

\usepackage{amsmath, amssymb}
\usepackage{amsmath}
\usepackage{bm}
\usepackage{graphicx} 
\usepackage{natbib}

\usepackage{url}
\usepackage{mathtools}
\newcommand{\bepsilon}{\boldsymbol{\epsilon}}

\newcommand{\mR}{\mathbb{R}}

\newcommand{\bY}{\mathbf{Y}}
\newcommand{\by}{\mathbf{y}}

\newcommand{\bc}{\mathbf{c}}
\newcommand{\bC}{\mathbf{C}}

\newcommand{\mE}{\mathbb{E}}

\newcommand{\bB}{\mathbf{B}}
\newcommand{\bU}{\mathbf{U}}

\newcommand{\KL}{\text{KL}}
\newcommand{\WassD}{\mathcal{W}_2}
\newcommand{\TV}{\text{TV}}
\newcommand{\intd}{\mathrm{d}}
\newcommand{\tarDist}{P_x}
\newcommand{\genDist}{Q_x}

\usepackage{algorithm}
\usepackage{algorithmic}
\usepackage{bm}
\usepackage{comment}
\usepackage{caption,subcaption}
\usepackage{float}
\usepackage{enumitem}

\newtheorem{theorem}{Theorem}

\newtheorem{lemma}{Lemma}
\theoremstyle{definition}
\newtheorem{assumption}{Assumption}
\theoremstyle{remark}

\title{Factor-Based Conditional Diffusion Model for Contextual Portfolio Optimization}
\author{Xuefeng Gao\thanks{Department of Systems Engineering and Engineering Management, The Chinese University of Hong Kong; xfgao@se.cuhk.edu.hk} \qquad 
Mengying He\thanks{Department of Systems Engineering and Engineering Management, The Chinese University of Hong Kong; myhe@se.cuhk.edu.hk} \qquad 
	Xuedong He\thanks{Department of Systems Engineering and Engineering Management, The Chinese University of Hong Kong; xdhe@se.cuhk.edu.hk} \qquad 
	Jiale Zha\thanks{Department of Systems Engineering and Engineering Management, The Chinese University of Hong Kong; jialezha@link.cuhk.edu.hk} }

\begin{document}
	\maketitle
	
	\begin{abstract}
		We propose a novel conditional diffusion model for contextual portfolio optimization that learns the cross-sectional distribution of next-day stock returns conditioned on high-dimensional asset-specific factors. Our model leverages a Diffusion Transformer architecture with token-wise conditioning, which enables linking each asset’s return to its own factor vector while capturing complex cross-asset dependencies. By drawing generative samples from the learned conditional return distribution, we perform daily mean–variance and mean-CVaR optimization, incorporating transaction costs and realistic constraints. Using data from the Chinese A-share market, we demonstrate that our approach consistently outperforms various standard benchmarks across multiple risk-adjusted performance metrics. Furthermore, we establish a 2-Wasserstein error bound for the conditional diffusion model and quantify how its distributional approximation errors propagate to the downstream portfolio optimization task. Our results demonstrate the potential of generative diffusion models for high-dimensional, risk-sensitive contextual stochastic optimization and financial decision making.
	\end{abstract}
	
	\section{Introduction}\label{sec:Intro}

Diffusion-based generative models are a powerful class of probabilistic generative AI models that can generate samples from high-dimensional target data distributions given a set of training data \citep{SongErmon2019, Ho2020}. 
The key idea is to use a forward process to gradually turn the unknown target distribution to a simple noise distribution, and then reverse this process to generate new samples. Recent empirical literature demonstrates that this iterative denoising mechanism, coupled with a stable training procedure, allows diffusion models to consistently outperform alternative generative frameworks—including generative adversarial networks \citep[GANs;][]{goodfellow2014generative}, variational autoencoders \citep[VAEs;][]{diederik2019introduction}, and normalizing flows \citep{kobyzev2020normalizing}—across diverse applications such as image generation, audio synthesis and computational biology \citep{dhariwal2021diffusion, rombach2022high, watson2023novo, yang2023diffusion}.


There are two fundamental paradigms for diffusion models: unconditional and conditional. While unconditional models are typically used to explore the upper limit of the performance of the generative model, conditional models focus on practical applications by generating samples tailored to specific input variables \citep{yang2023diffusion}. Given the remarkable success of conditional diffusion models in fields like text-to-image synthesis, a compelling question arises: how can these generative techniques be applied for financial decision-making such as portfolio optimization? This paper investigates this possibility within the framework of contextual (or conditional) portfolio optimization \citep{wang2022robust, nguyen2025robustifying}.

In contextual portfolio optimization, one aims to determine optimal asset allocations based on contextual or side information. Following the Arbitrage Pricing Theory \citep{ross2013arbitrage}, numerous studies exploit contextual information, such as macroeconomic factors and technical trading signals, to explain or predict the \textit{expected asset returns}. In this paper, we consider risk-averse contextual portfolio optimization, where the investor aims to minimize a risk function of the portfolio loss conditional on the context, by allocating weights to different stocks. In this case, the optimization objective is generally nonlinear in the distribution of stock returns conditioned on the context. To model this conditional distribution, 
we follow the general asset pricing framework of \citet{gu2020empirical}, and assume stock returns follow a statistical factor model. Specifically, the return vector is decomposed into {an unknown, potentially non-linear} function of observable factors (i.e. contexts) and a random shock with an unknown distribution. Such factor models are standard and widely adopted in the finance and econometrics literature \citep{kelly2023financial, chen2025diffusion}.

To solve the risk-averse contextual portfolio optimization problems in a data-driven manner, 
the primary challenge stems from learning the unknown conditional distribution of the stock return for a given factor, where both the return and the factors are continuous and potentially high-dimensional (e.g. on the order of hundreds), and the conditional return distribution is generally non-Gaussian. In particular, for any specific factor, the scarcity of historical return samples severely hinders direct estimation of this conditional distribution. 
This difficulty is further exacerbated by the high dimensionality of the return space when the number of {stocks} is large, and classical non-parametric density estimation methods struggle in such high-dimensional settings. To circumvent these issues, our key idea is to leverage conditional diffusion models as stochastic samplers to learn the conditional return distribution for any given factor.
By taking the factor as the condition or the input, the conditional diffusion model is able to output/generate a large set of synthetic return samples from the desired conditional distribution. These generated samples enable us to derive a sample version of the original stochastic contextual portfolio optimization problem, allowing us to derive approximate solutions. 

The main contributions of our paper can be summarized as follows.

\begin{itemize}

\item We propose to use a factor-based conditional diffusion model for solving contextual portfolio optimization. This approach enables the model to learn and generate the complete cross-sectional distribution of next-period stock returns, conditioned on asset-specific factors. Inspired by the success of Diffusion Transformer (DiT, \citet{peebles2023scalable}) architecture in image generation, we adapt the DiT architecture to incorporate a token-wise conditioning mechanism. This modification integrates asset-level factor information for each asset, while allowing the model to effectively capture complex dependencies between assets. The Transformer architecture is particularly well-suited for this task, due to its scalability, proficiency in modeling high-dimensional data, and capacity to accommodate diverse conditioning inputs.
 To our best knowledge, this is the first study to apply conditional diffusion models for risk-sensitive contextual (portfolio) optimization, highlighting the value of generative modeling for financial decision-making.

    \item We conduct an extensive empirical analysis on the Chinese A-share market to evaluate the performance of our proposed approach. The results demonstrate that mean-variance and mean-CVaR (Conditional Value-at-Risk) portfolios, optimized daily using our generated return distributions, significantly outperform those based on standard empirical and shrinkage estimators \citep{james1992estimation}, the equal-weighted strategy \citep{demiguel2009optimal}, and the DCC-GARCH model in \citep{engle2002dynamic}. Crucially, our analysis considers transaction costs in the construction and
evaluation of portfolio strategies, whereas these costs are ignored in some studies in the literature (e.g. \cite{nguyen2025robustifying}). 
More recently, \cite{bagchi2026factor} adopts our proposed approach in the context of the US stock market, further validating its effectiveness with encouraging empirical findings.

    \item We provide a rigorous theoretical error analysis that quantifies how the distributional approximation error of the conditional diffusion model propagates into downstream portfolio optimization. For both contextual mean-variance and mean-CVaR formulations, we derive explicit bounds for the suboptimality gap between the objective value of the true optimal solution and the solution obtained from diffusion-generated distributions. These bounds scales linearly with respect to the 2-Wasserstein distance (\(\mathcal{W}_{2}\)) between the two distributions.
Furthermore, we provide, to our knowledge, the first \(\mathcal{W}_{2}\) error bound for the \textit{conditional} diffusion model, directly linking this distributional gap to the model's training loss. Together, these bounds provide an end-to-end guarantee: as the conditional diffusion model minimizes its training loss, it more accurately captures the true conditional return distribution, ensuring that the resulting portfolio decisions converge toward optimality.

    

\end{itemize}

\subsection{Related Work}

The application of generative models to financial problems has 
grown rapidly in the past decade, driven by their ability to approximate complex, 
high-dimensional distributions that are difficult to characterize 
with classical parametric methods. For example, 
GANs have been adopted 
for tasks such as financial time series generation 
\citep{liao2024sig}, tail risk scenarios simulation \citep{cont2025tail}, limit order book 
simulation \citep{coletta2023conditional}, 
volatility surfaces modeling \citep{vuletic2024volgan}, data-driven hedging \citep{cont2025data}, and portfolio selection \citep{cetingoz2025synthetic}.
However, the adversarial training paradigm inherent in GANs presents significant practical challenges, most notably training instability and mode collapse.

Diffusion models have recently been extended to the financial domain, particularly for generating synthetic financial time series \citep{huang2024generative}. Nevertheless, leveraging these models for downstream decision-making tasks, such as portfolio optimization, remains a  relatively underexplored field. Recently, \citet{chen2025diffusion} proposes an \textit{unconditional} diffusion model with a \textit{latent (unobservable)} factor structure aimed primarily at dimensionality reduction in high-dimensional return modeling. They show that data generated by their unconditional diffusion model improves both mean and covariance estimation, leading to superior mean-variance optimal portfolios and factor portfolios. 
In contrast to \cite{chen2025diffusion}, we tackle a different problem with a different approach: we consider \textit{contextual} mean-variance and mean-CVaR portfolio optimizations and develop a \textit{conditional} diffusion model that leverages a rich set of \textit{observable} factors to generate/predict the next-period cross-sectional return distribution, which serve as a direct input for portfolio optimization. 
\citet{gao2026data} use conditional diffusion models to generate solution paths of stochastic differential equations, which are then used to enhance model-free reinforcement learning for continuous-time mean-variance portfolio selection. Meanwhile, \citet{aghapour2025solving} propose an approach for solving discrete-time dynamic portfolio selections by combining diffusion models for time series data generation with policy gradient algorithms.
In contrast to these studies, our work focuses on contextual portfolio optimization, utilizing diffusion models and various price-volume-based factors to directly learn high-dimensional conditional return distributions.

Our work is related to contextual stochastic optimization; see e.g. \cite{bertsimas2020predictive},
\cite{elmachtoub2022smart}, \cite{tao2025risk},
and the comprehensive survey by \cite{sadana2025survey}. Our approach for solving contextual portfolio optimization is known as Sequential Learning and Optimization (SLO), also referred to as predict-then-optimize. This approach first trains models to predict the conditional distribution of problem data based on observed context, then solves the resulting optimization problem. As discussed in Section 4.1 of \cite{sadana2025survey}, most recent literature in this domain typically assumes discrete conditional distributions. In contrast, our problem involves high-dimensional continuous return distributions conditioned on high-dimensional factors (i.e., context), and we leverage a conditional diffusion model to learn the conditional distribution of the next-period return, providing a computationally scalable solution. Once the return distribution has been learned, the downstream (sample-based) mean-variance and mean-CVaR portfolio optimizations can be easily solved via quadratic programs and linear programs, respectively. Recently, the working paper \cite{yoon2025data} combined Gaussian mixture models with traditional normalizing flows to capture multimodal conditional distributions in contextual stochastic optimization. However, the architectural constraints and high computational overhead of normalizing flows often limit their scalability and flexibility compared to diffusion models in generative tasks.
Finally, note that while one could adopt integrated learning and optimization (also known as end-to-end or decision-focused learning \cite{zhao2026diffusion, wang2026gen}) to train the predictive component specifically for downstream portfolio decisions, we opt out of this approach in this paper. The SLO approach we adopt offers distinct computational and flexible advantages in our setting by circumvention of retraining the diffusion model when investors evaluate alternative risk objectives.

Our theoretical error analysis is also broadly related to the literature on sensitivity analysis of portfolio 
optimization. 
\citet{best1991sensitivity} and \citet{chopra1993effect} show 
that mean--variance portfolio weights are extremely sensitive to 
perturbations in expected returns, and
\citet{lim2011conditional} analyze the 
fragility of sample-based CVaR portfolios. These analyses, however, operate at the parameter level, examining how errors in moments (means and covariances) or empirical tail loss propagate into portfolio weights—rather than measuring the discrepancy between the full joint return distributions. 
While the Wasserstein distributionally robust optimization literature \citep{mohajerin2018data, blanchet2022distributionally} does operate at the distributional level, its objective fundamentally differs from ours; notably, our analysis introduces no modifications to either the objective function or the constraints of the underlying contextual portoflio optimization problem.


In addition, our work extends the theoretical error bound literature on conditional diffusion models. Recent studies \citep{fu2024unveil, tangconditional} primarily investigate the statistical theory of these models, analyzing how effectively the conditional score function can be learned from finite training samples in score-based diffusion models \citep{songscore}. In contrast, we focus on the sampling theory by quantifying the \(\mathcal{W}_{2}\) convergence error of a well-trained conditional diffusion model. While several recent works have established \(\mathcal{W}_{2}\) error bounds for \textit{unconditional} diffusion models \citep[e.g.,][]{chen2023sampling, gao2025wasserstein, tang2025contractive, arsenyan2026assessing}, they typically rely on stringent assumptions—such as log-concave or bounded data distributions—which financial return data seldom satisfy. To circumvent these rigid requirements, we first derive an error bound in KL divergence for conditional diffusion models and subsequently convert it into a \(\mathcal{W}_{2}\) error bound between the true conditional return distribution and its generated counterpart, under realistic assumptions on the return distribution.

Finally, our paper is broadly related to studies on using machine learning for designing portfolio strategies. 
For example, \cite{ban2018machine} introduce performance-based
regularization to portfolio optimization by constraining the sample variances of the estimated portfolio risk and return. Their work does not consider contextual information and hence diverges significantly from ours. Additionally, \citep{gu2020empirical} and \citep{jiang2023re}  have applied deep learning approaches to predict future mean returns or the direction of future returns to build long-short portfolios. In contrast, our generative approach can learn the entire conditional return distribution, enabling the construction of more diverse portfolios for investors with different risk objectives including mean-variance and mean-CVaR. Furthermore, reinforcement learning (RL) has also been applied to data-driven portfolio optimization (e.g., \cite{wang2020continuous, hambly2023recent}). Compared to ``black-box'' RL policies, our approach decouples return prediction from portfolio optimization, which allows us to better incorporate factors (or side information) in portfolio decisions and adapt to non-stationary environments.

The remainder of the paper is organized as follows. Section~\ref{sec:problem} formulates the problem. Section~\ref{sec:ddpm} discusses learning return distribution via conditional diffusion models. Section~\ref{sec:empirical} evaluates the empirical effectiveness of our proposed generative framework for portfolio selection.
In Section~\ref{sec:sensitivity}, we provide a theoretical error analysis of portfolio optimization performance under the factor-based diffusion approach. The paper concludes in Section~\ref{sec:conclusion}. Supplementary details, robustness checks, and proofs are provided in the Appendix.


 \section{Problem Formulation}\label{sec:problem}

We follow the Arbitrage Pricing Theory \citep{ross2013arbitrage} and the general asset pricing framework used in \citet{gu2020empirical} to model stock returns using the following statistical factor model: 
\begin{equation}\label{eq:model}
    R_{t+1} = f(X_{t}) + u_{t+1},
\end{equation}
 where $R_{t+1} \in \mathbb{R}^{D}$ is the return vector of $D$ stocks in the period from time $t$ to $t+1$, $X_{t}= (x_{1,t}, x_{2,t}, ...., x_{D,t})^{\prime}\in \mathbb{R}^{D \times K}$ is a factor matrix observable at time $t$, $u_{t+1}$ is a (stationary) random shock independent of the information at time $t$, and $f$ is a deterministic function prescribing the dependence of the stock return on the factors. The form of $f$ and the distribution of $u_{t+1}$ are unknown, so the conditional distribution of $R_{t+1}$ given $X_t$ is also unknown and will be learned from data.

 Given the observable factor matrix $X_{t}=x$ at time $t$, the decision maker aims to solve the following contextual (or conditional) stochastic optimization problem: 
 \begin{equation}\label{eq:mean_risk}
\begin{aligned}
\max_{\bm{\omega} \in \mathcal{W}} \quad & -\mathrm{\rho}_{\mathbb{P}_{R_{t+1}|X_t=x}}(-\bm{\omega}^\top R_{t+1}) 
\end{aligned}
\end{equation}
where $\bm{\omega}=(\omega_1,\dots, \omega_D)'$ denotes the portfolio weight vector (the decision variable), $\mathcal{W}$ is the feasible region, 
$\mathbb{P}_{R_{t+1}|X_t=x}$ is the conditional distribution of the next period return given the current factor $X_t=x$,  $\mathrm{\rho}_{\mathbb{P}_{R_{t+1}|X_t=x}}$ is a risk function that maps the portfolio loss $-\bm{\omega}^\top R_{t+1}$ to a real number and captures the risk preference of the decision maker. A simple example of $\mathcal{W}$ is $\mathcal{W}=\{\bm{\omega}\in\mathbb{R}^{D}:
\mathbf{1}^\top\bm{\omega}=1\}$,
which imposes a fully-invested budget constraint. Two representative examples of the optimization problem \eqref{eq:mean_risk} are given below:

\noindent\textbf{Example 1 (Mean-Variance).}
Let $\rho_{\mathbb{P}_{R_{t+1}|X_t=x}}(-\bm{\omega}^\top R_{t+1}) 
= \mathbb{E}[-\bm{\omega}^\top R_{t+1} \mid X_t=x] 
+ \frac{\gamma}{2}\,\mathrm{Var}(-\bm{\omega}^\top R_{t+1} \mid X_t=x)$, 
where $\gamma > 0$ is the risk-aversion parameter. 
Then the objective in~\eqref{eq:mean_risk} becomes the (contextual) mean-variance problem of \citet{MarkowitzH:52ps}:
\begin{align*}
  \max_{\bm{\omega} \in \mathcal{W}} 
  \mathbb{E}\big[\bm{\omega}^\top R_{t+1} \mid X_t=x\big]
  -
  \frac{\gamma}{2}\,
  \mathrm{Var}\big(\bm{\omega}^\top R_{t+1} \mid X_t=x\big).
\end{align*}

\noindent\textbf{Example 2 (Mean-CVaR).}
Let $\rho_{\mathbb{P}_{R_{t+1}|X_t=x}}(-\bm{\omega}^\top R_{t+1})
= \mathbb{E}[-\bm{\omega}^\top R_{t+1} \mid X_t=x]
+ \frac{\Gamma}{2}\,\mathrm{CVaR}_{\beta}(-\bm{\omega}^\top R_{t+1} \mid X_t=x)$,
where $\Gamma > 0$ is the tail risk-aversion parameter and 
$\beta \in (0,1)$ is the confidence level. 
Here, CVaR is the Conditional Value-at-Risk, defined as:\footnote{A more rigorous definition of $\mathrm{CVaR}_\beta(-\bm{\omega}^\top R_{t+1})$ would be
    $\frac{1}{1-\beta}\int_{\beta}^1 \mathrm{VaR}_\alpha(-\bm{\omega}^\top R_{t+1})d\alpha.$
When $-\bm{\omega}^\top R_{t+1}$ has a continuous distribution, which is the case in our problem, this definition is the same as \eqref{eq:cvar}.}
\begin{equation}\label{eq:cvar}
\mathrm{CVaR}_\beta(-\bm{\omega}^\top R_{t+1}) = \mathbb{E}[-\bm{\omega}^\top R_{t+1} \mid -\bm{\omega}^\top R_{t+1} \geq \mathrm{VaR}_\beta(-\bm{\omega}^\top R_{t+1})],
\end{equation}
where $\mathrm{VaR}_\beta(-\bm{\omega}^\top R_{t+1}) := \inf\{l : \Pr(-\bm{\omega}^\top R_{t+1} \leq l) \geq \beta\}$.
The problem~\eqref{eq:mean_risk} {then} becomes the (contextual) mean-CVaR problem \citep{rockafellar2000optimization}:
\begin{align*}
  \max_{\bm{\omega} \in \mathcal{W}} 
  \mathbb{E}\big[\bm{\omega}^\top R_{t+1} \mid X_t=x\big]
  -
  \frac{\Gamma}{2}\,\mathrm{CVaR}_{\beta}\big(
    -\bm{\omega}^\top R_{t+1} \mid X_t=x\big).
\end{align*}

This paper focuses on the contextual mean-variance and mean-CVaR problems, and 
we also consider related portfolio optimization problems that take the transaction costs into account. In practice, rebalancing the portfolio incurs transaction costs. 
Let $\bm{\omega}^{d}$ denote the portfolio weight 
right before rebalancing, and let 
$c(\tau,\bm{\omega},\bm{\omega}^{d} )$ be a transaction 
cost function that quantifies the cost of rebalancing from 
$\bm{\omega}^{d}$ to a new target weight $\bm{\omega}$, where $\tau$ is the transaction cost parameter. 
The optimization problem~\eqref{eq:mean_risk} then becomes
\begin{equation}\label{eq:mean_risk_tc}
  \max_{\bm{\omega} \in \mathcal{W}} 
  -\rho_{\mathbb{P}_{R_{t+1}|X_t=x}}\big(-\bm{\omega}^\top 
  R_{t+1}\big)
  -
  c \big(\tau, \bm{\omega}, \bm{\omega}^{d}\big).
\end{equation}
A simple example of the transaction cost $c$ is the proportional transaction cost 
where $c(\tau, \bm{\omega},\bm{\omega}^{d})
= \sum_{i=1}^{D} \tau_i |\omega_{i}-\omega_{i}^{d}|$.

In practice, we have access to a set of data $(x_t,r_{t+1})_{t=0,\dots, T-1}$ as $T$ samples of $(X_t,R_{t+1})$ for some $T$. To solve the optimization problems \eqref{eq:mean_risk} and \eqref{eq:mean_risk_tc} in a data-driven manner, 
the primary challenge stems from learning the unknown conditional distribution $\mathbb{P}_{R_{t+1}|X_t=x}$ for each factor matrix $x \in \mathbb{R}^{D \times K}$, where both the return and the factor matrix are continuous and potentially high-dimensional. For any specific $x$, the scarcity of historical return samples severely hinders direct estimation of this conditional distribution, especially when the number of assets $D$ is large. In this paper, we address this issue by leveraging conditional diffusion models to learn this conditional distribution for an arbitrary factor matrix. This will be discussed in the next section. Once the conditional distribution of return is learned, one can generate an arbitrary number of synthetic return samples from this distribution, enabling us to derive sample/empirical versions of the original contextual portfolio optimization problems \eqref{eq:mean_risk} and \eqref{eq:mean_risk_tc}, and obtain approximate solutions.

	\section{Learning Return Distributions with Conditional Diffusion Models}\label{sec:ddpm}
    \label{method}

To learn the conditional distribution of \(R_{t+1}\) given \(X_{t}\), we adopt the conditional Denoising Diffusion Probabilistic Model (DDPM; \citealp{Ho2020}), a framework that has achieved significant success in high-dimensional generative tasks such as image synthesis. Formally, let $\bC$ denote a random vector (e.g., context). Conditional diffusion models aim to generate samples from an unknown (conditional) target distribution $p_{\text{target}}(\cdot | \bc)$ on $\mR^{D}$ given $\bC = \bc$
based on an observed set of training pairs $\{(\bc_i, \by_i)\}_{i=1}^T$, where $\bc_i$ and $\by_i$ are samples drawn from $\bC$ and $p_{\text{target}}(\cdot \vert{} \bc_i)$, respectively. In our framework, $\bc_i$ represents the factor matrix \(x_{i}\) at time \(i\), while $\by_i$ corresponds to the realized return \(r_{i+1}\). We next provide a brief overview of conditional diffusion models within our context, which generate synthetic return samples by first transforming raw return data into pure noise and then learning a reverse denoising process to recover the return distribution conditioned on specific factors.


Given a pair $(X_t,R_{t+1})$, we add Gaussian noise to $R_{t+1}$ using a forward (Markov) process:
 \begin{align}\label{eq:forward}
    R_{t+1}^{(0)}=R_{t+1},\quad  R_{t+1}^{(n)} = \sqrt{1 - \eta_n} R_{t+1}^{(n-1)} + \sqrt{\eta_n} \epsilon_n, \quad n=1,\dots N,
 \end{align}
 where $\epsilon_n$'s are i.i.d. $D$-dimensional standard Gaussian random vector and $\{\eta_n\}_{n=1}^N \in (0,1)$ is a predefined variance schedule. Denote $\zeta_n = 1- \eta_n$ and $\bar{\zeta}_n = \prod_{s=1}^n \zeta_s$. Then, $R_{t+1}^{(n)}$ admits the following representation in terms of $R_{t+1}^{(0)}$:
\begin{equation*}
    R_{t+1}^{(n)}
    =
    \sqrt{\bar{\zeta}_n}\, R_{t+1}^{(0)}
    +
    \sqrt{1-\bar{\zeta}_n}\,\bar{\epsilon}_n,
    \qquad
    \bar{\epsilon}_n \sim \mathcal{N}(0,I_D).
\end{equation*}
One can choose $\{\eta_n\}_{n=1}^N$ such that 
$\bar{\zeta}_N$ is close to 0 and the terminal distribution of $R_{t+1}^{(N)}$ is almost standard Gaussian. For instance, a linear variance schedule was used in \citealp{Ho2020} where \(\eta _{n}\) scales linearly from \(\eta_1 = 10^{-4}\) to \(\eta_N = 0.02\) across \(N=1000\) steps. One then gradually removes noise by running a learnable Markov process in the reverse
time direction. Specifically, the reverse denoising process is initialized with $\widetilde{R}_{t+1}^{(N)} \sim \mathcal{N}(0,I_D)$ (as an approximation to the true distribution of $R_{t+1}^{(N)}$). 
The learnable reverse transition kernel is then modeled as a Gaussian distribution:
we recursively draw $\widetilde{R}_{t+1}^{(n-1)}$ from 
 \begin{equation}\label{eq:reverse}
\mathcal{N}\left( \frac{1}{\sqrt{\zeta_n}}\left(\widetilde{R}^{(n)}_{t+1}-\frac{\eta_n}{\sqrt{1-\bar\zeta_n}}\bepsilon_\theta\left(\widetilde{R}^{(n)}_{t+1},n;X_{t}\right)\right), \sigma_n^2I_D\right), n=N,\dots, 1,
\end{equation}
where $\sigma_n^2:=\frac{1-\bar\zeta_{n-1}}{1-\bar\zeta_n}\eta_n$ and $\bepsilon_\theta\left(\cdot\right)$ is a neural network with parameter $\theta$ which will be trained using data. See \citealp{Ho2020} for details. The loss function for training is
\begin{align}\label{eq:loss}
    L(\theta):=\frac{1}{N}\sum_{n=1}^N\mathbb{E}\left[\|\epsilon-\bepsilon_\theta\left(R^{(n)}_{t+1},n;X_{t}\right)\|^2 \right],
\end{align}
where $\epsilon$ is a $D$-dimensional standard Gaussian random variable and the expectation is taken with respect to $(X_{t},R_{t+1})$ and $\epsilon$. 
One can interpret the loss \eqref{eq:loss} as the mean-squared error of noise prediction, where the neural network $\bepsilon_\theta$ is used to predict the noise $\epsilon$ added in the forward process given $\left(R^{(n)}_{t+1},n;X_{t}\right).$
This loss function can be estimated by using the data $(x_t,r_{t+1}),t=0,\dots, T-1$ as $T$ samples of $(X_{t},R_{t+1})$ and the samples of Gaussian noise $\epsilon$. The stochastic gradient descent algorithm and its variants can be applied to find (approximately) the optimal parameter value $\theta^*$. Once the model is trained, one can run the reverse process \eqref{eq:reverse} to generate new samples from the desired conditional distribution $\mathbb{P}_{R_{t+1}|X_t=x}$ for any given (potentially unseen) factor matrix $x$. 

\begin{figure}
  \centering
  \includegraphics[width=0.98\linewidth, height=10.5cm, keepaspectratio]{poster.png} 
  \caption{Modified DiT Architecture.}
  \label{fig:DiT}
\end{figure}

We adopt the Diffusion Transformer (DiT; \citet{peebles2023scalable}) 
architecture as the denoising network for $\bepsilon_\theta$, motivated by its scalability and its effectiveness in conditional image and video generation tasks. Below we first describe the overall 
architecture, then detail the internal structure of the DiT block, 
and finally highlight three key modifications that we introduce to adapt 
DiT to cross-sectional stock return modeling for portfolio optimization.

\textbf{Overall architecture.}
The overall structure is illustrated in Figure \ref{fig:DiT}. At diffusion step $n$, the model receives 
the noisy return vector $R_{t+1}^{(n)} \in \mathbb{R}^{D}$ together 
with the factor matrix $X_t \in \mathbb{R}^{D \times K}$ and the 
timestep index $n$. Each asset's scalar noisy return $R_{t+1,i}^{(n)}$ 
is first mapped to a $d_{\text{model}}$-dimensional return embedding 
via a linear projection, i.e., the scalar is multiplied by a 
learnable weight vector $\mathbf{w} \in \mathbb{R}^{d_{\text{model}}}$ and a learnable bias $\mathbf{b} \in \mathbb{R}^{d_{\text{model}}}$ is added. In the original DiT for image generation, an 
image is divided into non-overlapping patches and each patch is 
treated as a token in the input sequence. In our setting, each 
asset plays the role of a patch: every asset's return embedding 
constitutes one token, and the $D$ assets together form a sequence 
of length $D$. 
Simultaneously, the diffusion timestep 
$n$ is encoded into a vector $e_n \in \mathbb{R}^{d_{\text{model}}}$, 
while each asset's factor vector $x_{i,t} \in \mathbb{R}^K$ is 
processed by a (Factor) MLP (Multi-Layer 
Perceptron) to produce a factor embedding 
of the same dimension $d_{\text{model}}$. These two embeddings are summed element-wise to form a 
per-token condition vector 
$c_i = \mathrm{MLP}_{\mathrm{factor}}(x_{i,t}) + e_n$,\footnote{This procedure follows 
the original DiT architecture, where 
the class-label embedding and the timestep embedding are combined 
via element-wise addition.} which is 
then used to calculate the parameters that scale and shift the returns embeddings within each DiT block through another (condition) MLP. The 
sequence of $D$ returns embeddings, together with their corresponding 
condition vectors, is passed through a stack of $L$ identical DiT 
blocks. After the final DiT block, an Adaptive Layer Normalization (AdaLN, \citet{peebles2023scalable}) block is 
applied, which first normalizes each asset's 
$d_{\text{model}}$-dimensional return embedding to zero mean and 
unit variance, and then adaptively scales and shifts it using 
parameters generated from the condition vector $c_i$. This is followed by a linear layer, which is a single fully connected layer that projects each $d_{\text{model}}$-dimensional return embedding back to a scalar, producing the predicted noise vector 
$\bepsilon_\theta(R_{t+1}^{(n)}, n; X_t) \in \mathbb{R}^{D}$.

\textbf{DiT block.}
The right panel of Figure~\ref{fig:DiT} details the internal 
structure of a single DiT block, which consists of two sub-blocks 
connected by residual connections. In the first sub-block, the input 
returns embeddings $\{h_i\}_{i=1}^{D}$ are processed by a Layer 
Normalization \citep{ba2016layer} block whose output is then scaled and shifted by 
per-token parameters $(\gamma_{i,1}, \beta_{i,1})$ derived from the 
condition vector $c_i$. The resulting representations are fed into a 
Multi-Head Self-Attention (MHSA, \citet{vaswani2017attention}) layer, in which every asset embedding 
interacts with every other asset embedding so that each asset's 
representation can aggregate information from all $D$ assets, 
thereby capturing cross-asset dependencies.  The attention output for 
each token is further multiplied by a per-token gating scale 
$\alpha_{i,1}$ before being added back to the input via a residual 
connection. The second sub-block follows an analogous structure: a 
Layer Normalization with per-token parameters 
$(\gamma_{i,2}, \beta_{i,2})$, followed by a Pointwise 
Feedforward Network (FFN, \citet{vaswani2017attention}), another per-token gating scale $\alpha_{i,2}$, 
and a residual connection. Concretely, letting $h_i^{\ell}$ denote 
the embedding of asset $i$ at the input of the $\ell$-th DiT block, 
the computation proceeds as:
\begin{align*}
    &\hat{h}_i^{\ell} = h_i^{\ell} + 
        \alpha_{i,1}\cdot 
        \mathrm{MHSA}\Big(
            \gamma_{i,1} \odot \mathrm{LN}(h_i^{\ell}) 
            + \beta_{i,1}
        \Big), \\
    &h_i^{\ell+1} = \hat{h}_i^{\ell} + 
        \alpha_{i,2} \cdot 
        \mathrm{FFN}\Big(
            \gamma_{i,2} \odot \mathrm{LN}(\hat{h}_i^{\ell}) 
            + \beta_{i,2}
        \Big),
\end{align*}
where $\mathrm{LN}(\cdot)$ denotes Layer Normalization, $\odot$ is 
the element-wise product, and all six parameters 
$(\gamma_{i,1}, \beta_{i,1}, \alpha_{i,1}, 
\gamma_{i,2}, \beta_{i,2}, \alpha_{i,2})$ are produced by a 
shared MLP from the condition vector $c_i$:
\begin{align*}
    \big(\gamma_{i,1}, \beta_{i,1}, \alpha_{i,1}, 
\gamma_{i,2}, \beta_{i,2}, \alpha_{i,2}\big) 
    = \mathrm{MLP}(c_i).
\end{align*}
Following \citet{peebles2023scalable}, all six scale, shift and gating parameters are initialized to zero.

\textbf{Key modifications.}
We introduce three modifications to adapt the original DiT to our cross-sectional return modeling:
\begin{itemize}[leftmargin=2em]
\item[1.] \textbf{Raw data modelling.} The original DiT operates on latent 
image patches produced by a Variational Autoencoder (VAE) 
. Since stock returns are 
low-dimensional compared with images, we eliminate the VAE and let each token directly 
represent a single asset's noisy return, mapped to the 
embedding space by linear transformation.
\item[2.] \textbf{Token-wise conditioning.} The original DiT applies 
a single global condition (e.g., a class label) identically to 
all tokens (image patches). In our 
setting, each asset possesses a distinct factors vector, so 
we assign each token its own condition 
$c_i = \mathrm{MLP}_{\mathrm{factor}}(x_{i,t}) + e_n$. Each 
asset's unique factor information is therefore captured through 
per-token conditioning, while cross-asset dependencies are modeled 
by MHSA, which allows every token to exchange information with all 
other tokens.
\item[3.] \textbf{Per-token AdaLN-Zero.} Correspondingly, the AdaLN 
scale, shift, and gating parameters are derived independently for 
each token from its own condition $c_i$, rather than from a shared global 
vector. This enables two assets with different factors to
receive different scale, shift, and gating parameters, and therefore undergo different transformations in every DiT block.
\end{itemize}

	\section{Empirical Evaluations and Results}\label{sec:empirical}

 In this section, we empirically demonstrate the effectiveness of our proposed factor-based conditional diffusion model in mean-variance portfolio selection and mean-CVaR portfolio selection  on the Chinese A-share market.

\subsection{Data and Experimental Setup}
We consider daily investment and use daily stock return data for the CSI 300 Index constituents in the period 4-Jan-2017 (time 1) to 9-Apr-2025, obtained from the Wind Database (\url{https://www.wind.com.cn/}). The corresponding stock-level factor data covering the period 3-Jan-2017 to 8-Apr-2025 is sourced from another leading financial data vendor Datayes (\url{https://www.datayes.com/}). We consider price-volume-based factors, resulting in 208 factors per stock. Representative examples of factors include moving averages of price (5/10/20/60-day), historical returns (5/10/20-day), return variance (20/60/120-day), exponential moving averages of volume (5/10/12/26-day), and moving averages of turnover rate (5/10/20/60/120-day).

The return and factor dataset is split chronologically into a training period from 4-Jan-2017 to 29-Dec-2023 and a test period from 2-Jan-2024 to 9-Apr-2025, maintaining an 8:2 train/test ratio. To ensure data consistency and avoid estimation artifacts from intermittent trading, we focus on a core universe of 118 liquid CSI 300 constituent stocks that have complete daily observations throughout the training period.
 For each day during the training period, factor values are standardized cross-sectionally (i.e., across all stocks) and winsorized at three standard deviations to mitigate the influence of outliers, with missing values imputed by the cross-sectional mean of that factor. Stock returns are winsorized similarly. The conditional diffusion model is trained using the Adam optimizer on an NVIDIA A30 GPU and the hyperparameters are optimized via grid search. The total training time is approximately 20 minutes.


During the test period, we perform daily portfolio rebalancing: on each trading day $t$, the trained diffusion model generates synthetic return samples conditioned on the current factor matrix $X_t$, from which the mean and covariance (or CVaR) are estimated and used to solve the portfolio optimization problem, and the resulting optimal weights are implemented for the next trading day. 
     Since our diffusion model generates the return vector of all 
$D=118$ stocks simultaneously, it requires a complete factor matrix 
$X_t$ as input to generate samples of day $t+1$ on every trading day. For stock factors that are missing because of stock suspension during the test period, we fill them with those 
from the most recent non-suspended trading day. While the model 
produces synthetic return samples for all 118 stocks, suspended 
stocks and stocks hitting their daily price limits are excluded 
from the portfolio optimization: their amounts are held fixed at 
their pre-rebalancing levels, and only the 
remaining stocks participate in the optimization over the residual 
portfolio weight. In addition, during the test period, Guotai Junan Securities
(stock code: 601211) merged with Haitong Securities (stock code: 600837). 
During Haitong Securities' brief resumption of trading prior to 
its delisting, we set its portfolio weight to zero to avoid 
holding a stock that is about to be delisted.

\subsection{Contextual Mean-variance portfolio optimization}
    In this section, we consider mean-variance portfolio optimization. 
We first consider the case where transaction costs are not taken into account in the optimization formulation. 
 Following the mean-variance formulation in Example~1 
    of Section~\ref{sec:problem}, the portfolio optimization problem at day $t$ becomes:
\begin{equation}\label{eq:mean_variance}
\begin{aligned}
\max_{\bm{\omega}} \quad & \bm{\omega}^\top \boldsymbol{\mu_{t+1}} - \frac{\gamma}{2}\bm{\omega}^\top \boldsymbol{\Sigma_{t+1}}\bm{\omega} \quad \\
\text{s.t.} \quad & \bm{\omega}^\top \mathbf{1} = 1, \quad \omega_i \geq 0, \quad \forall i,
\end{aligned}
\end{equation}
where $\bm{\omega}=(\omega_1,\dots, \omega_D)'$, $\gamma$, $\bm{\mu_{t+1}}$, and $\bm{\Sigma_{t+1}}$ denote the portfolio weight vector, the investor's risk aversion degree, the mean and covariance matrix of the stock return vector on day $t+1$ (conditional on the factor $X_t$), respectively, and the constraint $\omega_i\ge 0$ is imposed because short sales are not allowed in the A-share market.

We set $\gamma=100$ so as to generate a reasonable amount of risk taking. Robustness checks using alternative risk aversion coefficients are presented in Appendix~\ref{sec:robust}.
At each time $t$, we estimate $\bm{\mu_{t+1}}$ and $\bm{\Sigma_{t+1}}$, compute 
the optimal portfolio for the coming day, and implement the portfolio. 
We use four estimation methods: (i) Factordiff, which estimates the 
mean and covariance from predictive samples generated by our 
conditional diffusion model (we report results for 500, 1000, and 
2000 samples); (ii) Emp, which computes the sample mean 
$\bm{\bar{\mu}}_t$ and sample covariance $\bm{\bar{\Sigma}}_t$ using 
historical data from time 1 to $t$; (iii) ShrEmp, the James--Stein 
shrinkage estimator \citep{james1992estimation}; and (iv) DCC-GARCH (Dynamic Conditional Correlation--Generalized Autoregressive Conditional Heteroskedasticity), which computes the mean and the covariance from a fitted DCC-GARCH(1,1) model with Student-$t$ innovations 
\citep{bollerslev1986generalized, engle2002dynamic}. DCC-GARCH is a widely adopted multivariate time series 
model for capturing time-varying volatilities and correlations 
of financial returns. The parameters of the DCC-GARCH model are calibrated to the training data and fixed in the test period. We compare the performance of these four 
estimation methods as well as the equally weighted portfolio 
(EW; \citealp{demiguel2009optimal}) in terms of the mean, 
standard deviation, Sharpe ratio, Sortino ratio, Calmar ratio, and 
Return-to-CVaR (RtC) \citep{huang2024mean}. The precise definitions of these performance metrics are provided in Appendix~\ref{app:experiment_setting}. The results are shown in Table~\ref{without cost}. Factordiff and 
DCC-GARCH yield notably better risk-adjusted performance than EW, 
Emp, and ShrEmp when transaction costs are not taken into account. 
Among all methods, Factordiff~(500) achieves the highest mean return 
(0.110\%), Sharpe ratio (0.099), and Calmar ratio (0.011), while 
maintaining a competitive Sortino ratio and Return-to-CVaR compared 
with DCC-GARCH.

\begin{table}
  \caption{Performance of the EW portfolio and the optimal portfolio 
  of \eqref{eq:mean_variance} (with $\gamma=100$) with Factordiff 
  (500, 1000, and 2000 samples), Emp, ShrEmp, and DCC-GARCH estimates 
  of stock return moments. Transaction fees are ignored.}
  \label{without cost}
  \centering
  \small
  \begin{tabular}{l|cccccc}
    \toprule
    Method & Mean (\%) & Std (\%) & Sharpe ratio & Sortino ratio & Calmar ratio& RtC \\
    \midrule
    EW                 & 0.044             & 1.350              & 0.032             & 0.049             & 0.003             & 0.016             \\
    Factordiff (500)   & $\bm{0.110}$   & 1.115           & $\bm{0.099}$   & 0.146           & $\bm{0.011}$   & 0.044           \\
    Factordiff (1000)  & 0.104           & 1.116           & 0.094           & 0.140           & 0.011           & 0.041           \\
    Factordiff (2000)  & 0.103           & 1.114           & 0.092           & 0.136           & 0.011           & 0.040           \\
    Emp                & 0.074           & $\bm{0.948}$   & 0.078           & 0.115           & 0.008           & 0.034           \\
    ShrEmp             & 0.077           & 0.960           & 0.081           & 0.119           & 0.008           & 0.036           \\
    DCC-GARCH          & 0.089           & 0.968           & 0.092           & $\bm{0.150}$   & 0.008           & $\bm{0.046}$   \\
    \bottomrule
  \end{tabular}
\end{table}

\begin{figure}
  \centering
  \includegraphics[width=1\linewidth, height=12cm]{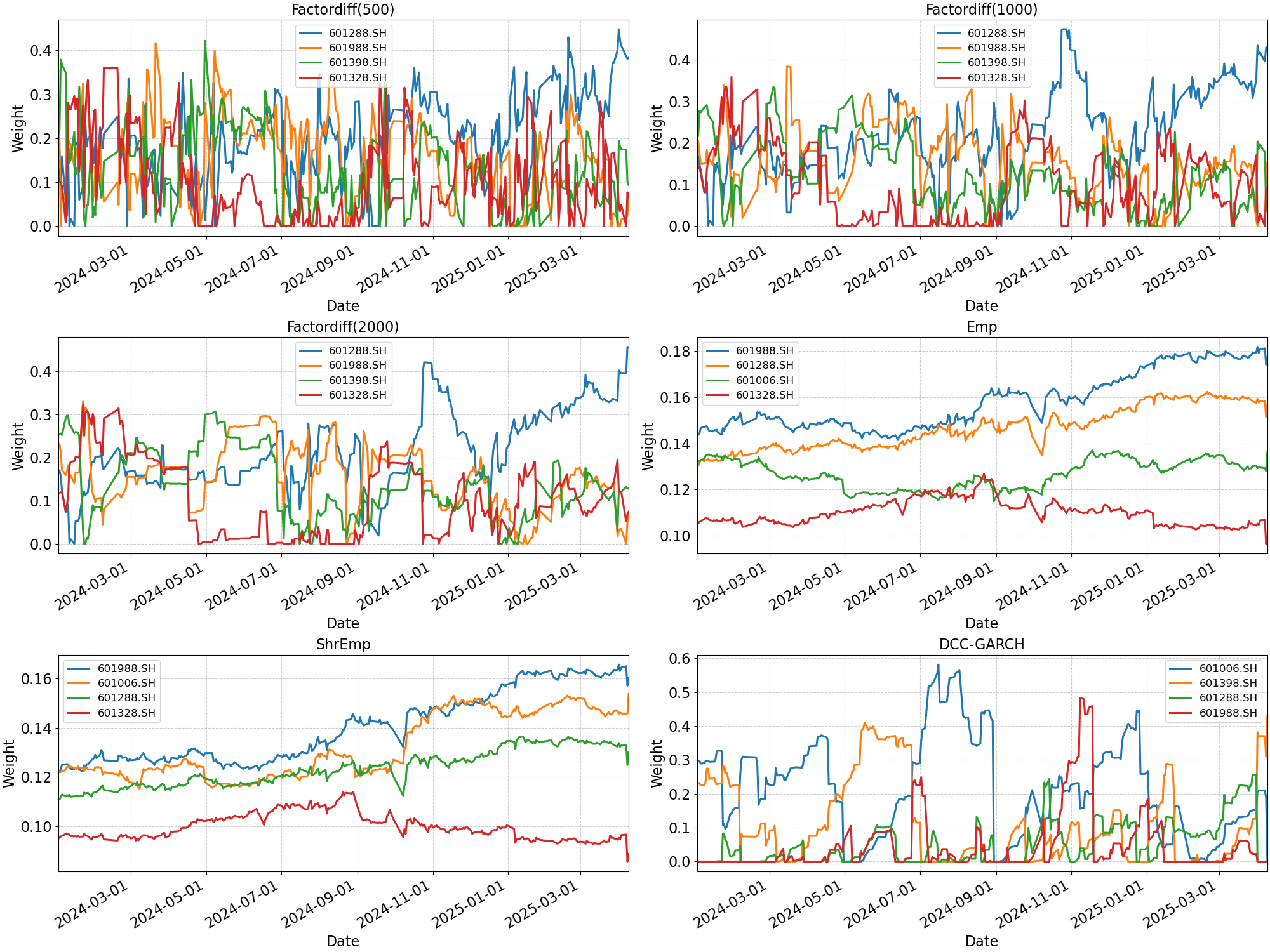} 
  \caption{Portfolio weights over time for the top 4 stocks in the optimal portfolio of \eqref{eq:mean_variance} with stock return moments estimated by Factordiff (500, 1000, and 2000 samples), Emp, ShrEmp, and DCC-GARCH.}
  \label{fig:weight_mv}
\end{figure}

In practice, 
trading incurs transaction costs, which arise because 
the portfolio weights are re-optimized on each trading day based on updated return estimates, and the resulting daily rebalancing incurs costs.
In the A-share market, 
the transcation fees include trading commissions, stamp tax (applied only to the 
seller), and slippage, which amount to approximately 7.5 basis points 
(bps) for buying orders and 12.5 bps for selling orders per unit 
trading amount \citep{leippold2022machine}. Table~\ref{with cost} 
reports the mean return, standard deviation, Sharpe ratio, Sortino 
ratio, Calmar ratio, and Return-to-CVaR of each portfolio with 
transaction fees deducted under problem \eqref{eq:mean_variance}. There are two observations. First, the 
transaction fees are negligible for the EW, Emp, and ShrEmp 
portfolios. Note that the portfolio weights are constant in EW and 
nearly constant in Emp and ShrEmp because a single new data point has 
negligible impact on the moment estimates given a large amount of 
existing data and, consequently, the day-by-day update of the 
empirical mean and covariance of stock returns is minimal. The daily 
stock price change is small, e.g., within 6\%, so the transaction 
fees due to portfolio rebalancing in the case of maintaining constant 
portfolio weights over time are minimal. Second, with transaction fees deducted, both Factordiff and 
DCC-GARCH underperform Emp and ShrEmp, which is due to the large 
amount of transaction fees that the former two strategies incur. This is because their 
portfolio weights vary significantly over time---Factordiff reacts 
to daily changes in the generated return distribution, and DCC-GARCH 
updates its volatility and correlation estimates at every time 
step---as shown in Figure~\ref{fig:weight_mv}. Moreover, as can be 
seen from the figure, the portfolio weights of Factordiff become 
smoother as the number of generated samples increases: 
Factordiff~(500) exhibits the most volatile weights and thus incurs 
the highest transaction fees, causing its mean return to drop from 
0.110\% to 0.052\% (a reduction of 0.058\% points), whereas 
Factordiff~(2000) experiences a much smaller reduction from 0.103\% 
to 0.081\% (0.022\% points). This is because a larger 
sample size yields more stable moment estimates and reduces day-to-day moments fluctuation, which in turn 
leads to lower portfolio turnover. As a result, Factordiff~(2000) 
achieves performance comparable to Emp and ShrEmp even after 
transaction fees are deducted.

\begin{table}
  \caption{Performance of the EW portfolio and the optimal portfolio 
  of \eqref{eq:mean_variance} (with $\gamma=100$) with Factordiff 
  (500, 1000, and 2000 samples), Emp, ShrEmp, and DCC-GARCH estimates 
  of stock return moments. Transaction fees are deducted.}
  \label{with cost}
  \centering
  \small
  \begin{tabular}{l|cccccc}
    \toprule
    Method & Mean (\%) & Std (\%) & Sharpe ratio & Sortino ratio& Calmar ratio& RtC \\
    \midrule
    EW                 & 0.043              & 1.350              & 0.032              & 0.048              & 0.003            & 0.015              \\
    Factordiff (500)   & 0.052           & 1.114           & 0.047           & 0.067           & 0.005           & 0.020           \\
    Factordiff (1000)  & 0.067           & 1.116           & 0.060           & 0.088           & 0.006           & 0.026           \\
    Factordiff (2000)  & $\bm{0.081}$   & 1.114           & 0.073           & 0.106           & $\bm{0.008}$   & 0.032           \\
    Emp                & 0.073           & $\bm{0.948}$   & 0.077           & 0.114           & 0.008           & 0.034           \\
    ShrEmp             & 0.077           & 0.960           & $\bm{0.080}$   & $\bm{0.118}$   & 0.008           & $\bm{0.035}$   \\
    DCC-GARCH          & 0.071           & 0.965           & 0.073           & 0.116           & 0.006           & 0.035           \\
    \bottomrule
  \end{tabular}
\end{table}

\begin{figure}[!htbp]
  \centering
  \includegraphics[width=1\linewidth, height=12cm]{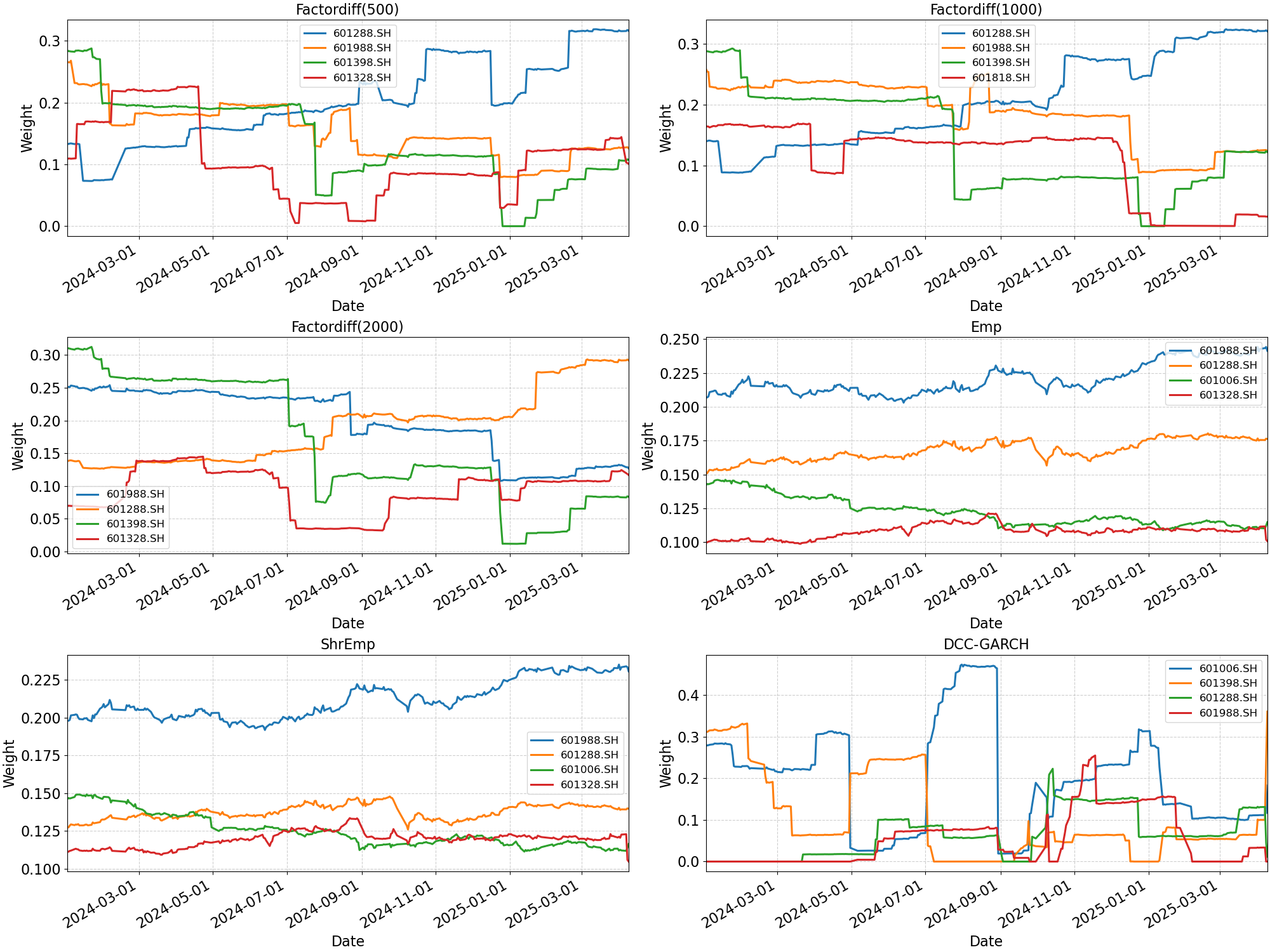} 
  \caption{Portfolio weights over time for the top 4 stocks in the optimal portfolio of \eqref{eq:mean_variance_tc} with stock return moments estimated by Factordiff (500, 1000, and 2000 samples), Emp, ShrEmp, and DCC-GARCH.}
  \label{fig:weight_mv_l1}
\end{figure}

The above findings show the importance of considering transaction 
fees in the construction and evaluation of portfolio strategies, 
whereas these fees are ignored in some studies in the literature. Hence, in the following we consider the problem of class \eqref{eq:mean_risk_tc}, where transaction costs are explicitly incorporated in the optimization formulation. Specifically, to 
account for transaction fees, we consider the optimization problem:
\begin{equation}\label{eq:mean_variance_tc}
\begin{aligned}
\max_{\bm{\omega}, \bm{b}, \bm{s}} \quad & \bm{\omega}^\top 
\boldsymbol{\mu_{t+1}} - \frac{\gamma}{2} \bm{\omega}^\top 
\boldsymbol{\Sigma_{t+1}} \bm{\omega} - \left(0.00075\, 
\bm{b}^\top \mathbf{1} + 0.00125\, \bm{s}^\top \mathbf{1} 
\right) \\
\text{s.t.} \quad & \bm{\omega}^\top \mathbf{1} = 1, 
 0\leq \omega_{i} \leq 1, \\
 & \bm{b} \geq 0, \bm{s} \geq 0,\\
&\omega_{i} - \omega^{d}_{i} = b_{i} - s_{i}, \quad \forall i,
\end{aligned}
\end{equation}
where $\bm{b}$ and $\bm{s}$ stand for the vectors of buying and 
selling amounts, per unit wealth, of the stocks due to portfolio 
rebalancing and $\omega^{d}_{i}$ denotes the dollar 
amount of stock $i$ right before portfolio 
rebalancing. Table~\ref{mv-l1} presents the performance of the EW 
portfolio and the optimal portfolio of \eqref{eq:mean_variance_tc} 
with estimates of the mean and covariance of stock returns based on 
Factordiff (500, 1000, and 2000 samples generated), Emp, ShrEmp, and 
DCC-GARCH, taking into account transaction fees directly in the 
optimization objective and transaction fees are deducted in the performance evaluation. We can see that Factordiff consistently 
outperforms the other strategies across all risk-adjusted performance 
metrics. Moreover, incorporating transaction fees directly into the optimization objective plays a crucial role in stabilizing Factordiff's performance across different sample sizes. When transaction fees are ignored in the objective function, Factordiff is sensitive to sample size—Factordiff (500) achieves the best Sharpe ratio in Table~\ref{without cost} but suffers the sharpest decline once fees are deducted in Table~\ref{with cost}, as fewer samples yield noisier estimates and hence higher turnover. In contrast, under \eqref{eq:mean_variance_tc}, the transaction cost penalty regularizes portfolio weights, and Factordiff's performance remains stable from 500 to 2000 samples (e.g., Sharpe ratio ranging from 0.090 to 0.092). We 
can also observe from Figure~\ref{fig:weight_mv_l1} that the Factordiff portfolio in 
\eqref{eq:mean_variance_tc} entails much smoother portfolio weights 
over time than the Factordiff portfolio in \eqref{eq:mean_variance}, 
so the former incurs much lower transaction fees, explaining its 
superior performance.

\begin{table}
  \caption{Performance of the EW portfolio and the optimal portfolio 
  of \eqref{eq:mean_variance_tc} (with $\gamma=100$) with Factordiff 
  (500, 1000, and 2000 samples), Emp, ShrEmp, and DCC-GARCH estimates 
  of stock return moments. Transaction fees are incorporated into the 
  optimization objective and deducted.}
  \label{mv-l1}
  \centering
  \small
  \begin{tabular}{l|cccccc}
    \toprule
    Method & Mean (\%) & Std (\%) & Sharpe ratio& Sortino ratio& Calmar ratio& RtC \\
    \midrule
    EW                 & 0.043              & 1.350              & 0.032              & 0.048              & 0.003            & 0.015              \\
    Factordiff (500)   & 0.100           & 1.109           & 0.090           & 0.132           & 0.010   & 0.039   \\
    Factordiff (1000)  & 0.100           & 1.102           & 0.090           & 0.132           & 0.010           & 0.039           \\
    Factordiff (2000)  & $\bm{0.102}$   & 1.110           & $\bm{0.092}$   & $\bm{0.134}$   & $\bm{0.010}$          & $\bm{0.039}$           \\
    Emp                & 0.072           & $\bm{0.968}$   & 0.075           & 0.109           & 0.008           & 0.032           \\
    ShrEmp             & 0.077           & 0.981           & 0.079           & 0.114           & 0.008           & 0.034           \\
    DCC-GARCH          & 0.076           & 0.992           & 0.077           & 0.120           & 0.006           & 0.036           \\
    \bottomrule
  \end{tabular}
\end{table}

\subsection{Contextual Mean-CVaR portfolio optimization}

We next consider the mean-CVaR portfolio optimization problem. Given the observed factor matrix $X_t$ at time $t$, the contextual mean-CVaR portfolio selection problem, without incorporating transaction costs into the objective, is formulated as:
\begin{equation}\label{eq:mean_CVaR}
\begin{aligned}
\max_{\bm{\omega}} \quad & \bm{\omega}^\top \boldsymbol{\mu_{t+1}} 
  - \frac{\Gamma}{2}\,\mathrm{CVaR}_\beta(-\bm{\omega}^\top R_{t+1}) \\
\text{s.t.} \quad & \bm{\omega}^\top \mathbf{1} = 1, \quad 
  \omega_i \geq 0, \quad \forall\, i,
\end{aligned}
\end{equation}
where $\beta$ is the confidence level and $\Gamma$ is the tail risk aversion coefficient. By the well-known 
variational representation of CVaR 
\citep{rockafellar2000optimization}:
\begin{equation}\label{eq:CVaR_re}
\begin{aligned}
 \text{CVaR}_\beta(-\bm{\omega}^\top R) = \inf_{\alpha \in \mathbb{R}} \left\{ \alpha + \frac{1}{1-\beta} E[(-\bm{\omega}^\top R - \alpha)^+] \right\},
\end{aligned}
\end{equation}
then the optimization problem \eqref{eq:mean_CVaR} can be reformulated as:
\begin{equation}\label{eq:mean_CVaR_mid}
\begin{aligned}
 \max_{\bm{\omega},\alpha} & \quad \bm{\omega}^\top \boldsymbol{\mu_{t+1}} - \frac{\Gamma}{2}\left\{ \alpha + \frac{1}{1-\beta} E[(-\bm{\omega}^\top R_{t+1} - \alpha)^+] \right\} \\
 \text{s.t.} & \quad \bm{\omega}^\top \mathbf{1} = 1, \
 \omega_i \geq 0, \quad \forall i,
\end{aligned}
\end{equation}
When the expectation is estimated by a sample average over multiple return 
samples, we can rewrite \eqref{eq:mean_CVaR_mid} as:
\begin{equation}\label{eq:mean_CVaR_lq}
\begin{aligned}
\max_{\bm{\omega},\, \alpha,\, \bm{z}} \quad & \bm{\omega}^\top \boldsymbol{\mu_{t+1}} - \frac{\Gamma}{2} \left( \alpha + \frac{1}{M(1 - \beta)} \sum_{j=1}^{M} z_j \right)  \\
\text{s.t.} \quad & \sum_{i} w_i = 1, \\
& w_i \geq 0 \quad \forall i, \\
& z_j + \bm{\omega}^\top r_{t+1,j} + \alpha \geq 0, \quad \forall j ,\\
& z_j \geq 0, \quad \forall j,
\end{aligned}
\end{equation}
where $M$ is the number of return samples $r_{t+1,j}$ obtained from the conditional return distribution (given $X_t$). This a linear program problem and can be solved efficiently.

We set $\Gamma = 1$ and $\beta=0.95$. {Robustness checks using alternative risk aversion coefficients are presented in Appendix~\ref{sec:robust}.} 
At each time $t$, we 
estimate $\boldsymbol{\mu}_{t+1}$ and generate return samples 
$\{r_{t+1,j}\}_{j=1}^{M}$, solve the portfolio optimization 
problem~\eqref{eq:mean_CVaR_lq} for the coming day, and implement 
the portfolio. We use four estimation methods: (i) Factordiff, which estimates $\boldsymbol{\mu}_{t+1}$ by the 
sample mean of predictive samples generated by our conditional 
diffusion model and directly uses those samples as the return 
samples $r_{t+1,j}$ (we report results for $M = 500$, $1000$, 
and $2000$); (ii) Emp, which estimates $\boldsymbol{\mu}_{t+1}$ by the sample 
mean $\bar{\boldsymbol{\mu}}_t$ using historical data from time $1$ 
to $t$ and uses the historical returns as the return samples 
$r_{t+1,j}$; (iii) ShrEmp, which estimates $\boldsymbol{\mu}_{t+1}$ by the 
James-Stein shrinkage estimator and uses 
the historical returns as the return samples $r_{t+1,j}$; 
and (iv) DCC-GARCH, which computes the mean and the CVaR from a fitted 
DCC-GARCH(1,1) model with Student-$t$ innovations. We evaluate all methods, along with the EW benchmark, using the same set of performance metrics as before.

Table~\ref{cvar_without_cost} reports the results when transaction costs are ignored in both the optimization formulation \eqref{eq:mean_CVaR_lq} and the empirical performance evaluation of various methods. We observe that Factordiff yields the best performance across all risk-adjusted metrics. In particular, Factordiff~(500) achieves the highest mean return (0.136\%), Sharpe ratio (0.126), Sortino ratio (0.197), Calmar ratio (0.017), and RtC (0.059), substantially outperforming Emp, ShrEmp, and DCC-GARCH. Among the competing methods, DCC-GARCH comes closest to Factordiff with a Sharpe ratio of 0.091, but still falls considerably short. 

\begin{table}[H]
  \caption{Performance of the EW portfolio and the optimal portfolio 
  of \eqref{eq:mean_CVaR_lq} (with $\Gamma=1$) with Factordiff 
  (500, 1000, and 2000 samples), Emp, ShrEmp, and DCC-GARCH estimates 
  of stock return distribution. Transaction fees are ignored.}
  \label{cvar_without_cost}
  \centering
  \small
  \begin{tabular}{l|cccccc}
    \toprule
    Method & Mean (\%) & Std (\%) & Sharpe ratio & Sortino ratio 
           & Calmar ratio & RtC \\
    \midrule
    EW                 & 0.044             & 1.350              & 0.032             & 0.049             & 0.003             & 0.016             \\
    Factordiff (500)   & $\bm{0.136}$   & 1.082           & $\bm{0.126}$   
                       & $\bm{0.197}$   & $\bm{0.017}$          
                       & $\bm{0.059}$   \\
    Factordiff (1000)  & 0.133           & 1.117           & 0.119           
                       & 0.186           & 0.017                  
                       & 0.055           \\
    Factordiff (2000)  & 0.122           & 1.114           & 0.109           
                       & 0.166           & 0.014                  
                       & 0.049           \\
    Emp                & 0.083           & $\bm{0.974}$   & 0.085           
                       & 0.126           & 0.009            
                       & 0.038           \\
    ShrEmp             & 0.079           & 0.985           & 0.080           
                       & 0.118           & 0.008                 
                       & 0.036           \\
    DCC-GARCH          & 0.094           & 1.036           & 0.091          
                       & 0.144           & 0.012                  
                       & 0.045           \\
    \bottomrule
  \end{tabular}
\end{table}

\begin{figure}[!htbp]
  \centering
  \includegraphics[width=1\linewidth, height=12cm]{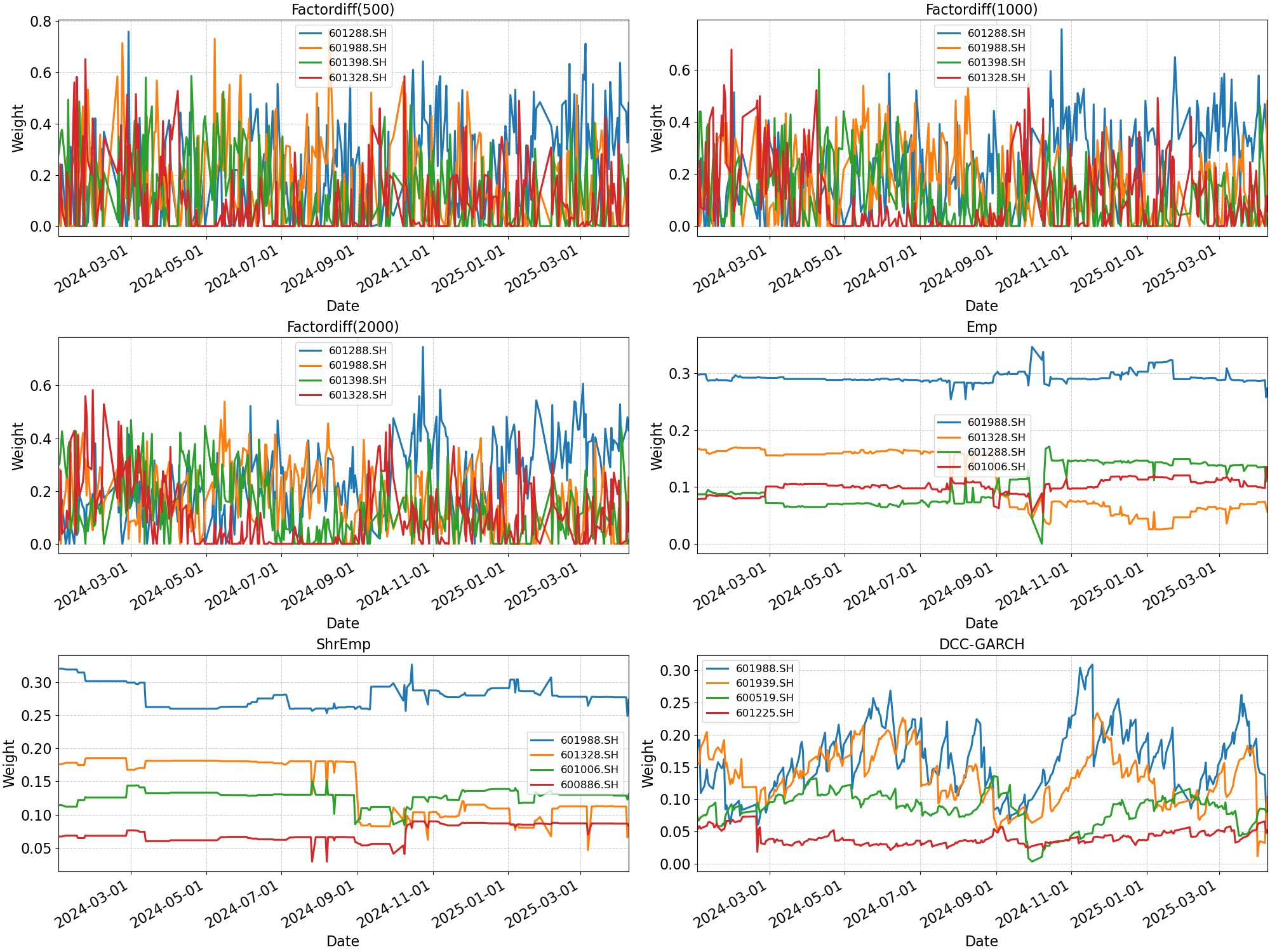} 
  \caption{Portfolio weights over time for the top 4 stocks in the optimal portfolio of \eqref{eq:mean_CVaR_lq} with stock return scenarios from Factordiff (500, 1000, and 2000 samples), Emp, ShrEmp, and DCC-GARCH.}
  \label{fig:weight_cvar}
\end{figure}

Table~\ref{cvar_with_cost} reports the results with transaction fees deducted in the empirical evaluation of different methods. As shown in Figure~\ref{fig:weight_cvar}, the portfolio weights of Factordiff fluctuate far more aggressively than those of the other methods, while DCC-GARCH exhibits moderate turnover—noticeably higher than Emp and ShrEmp, whose weights remain largely stable, but much lower than Factordiff. Consistent with our earlier analysis, Factordiff (500), which achieves the best performance when transaction fees are ignored, suffers the most severe degradation after fees are deducted (Sharpe ratio dropping from 0.126 to 0.020), and the degradation diminishes as the sample size increases (Factordiff (2000) drops from 0.109 to 0.042), confirming that fewer samples induce noisier estimates and hence higher turnover costs. The performance of Emp and ShrEmp is barely affected by transaction fees (e.g., Emp's Sharpe ratio drops only from 0.085 to 0.082), reflecting their low turnover. As a result, Factordiff underperforms the competing methods after transaction fees are deducted. 

\begin{figure}[!htbp]
  \centering
  \includegraphics[width=1\linewidth, height=12cm]{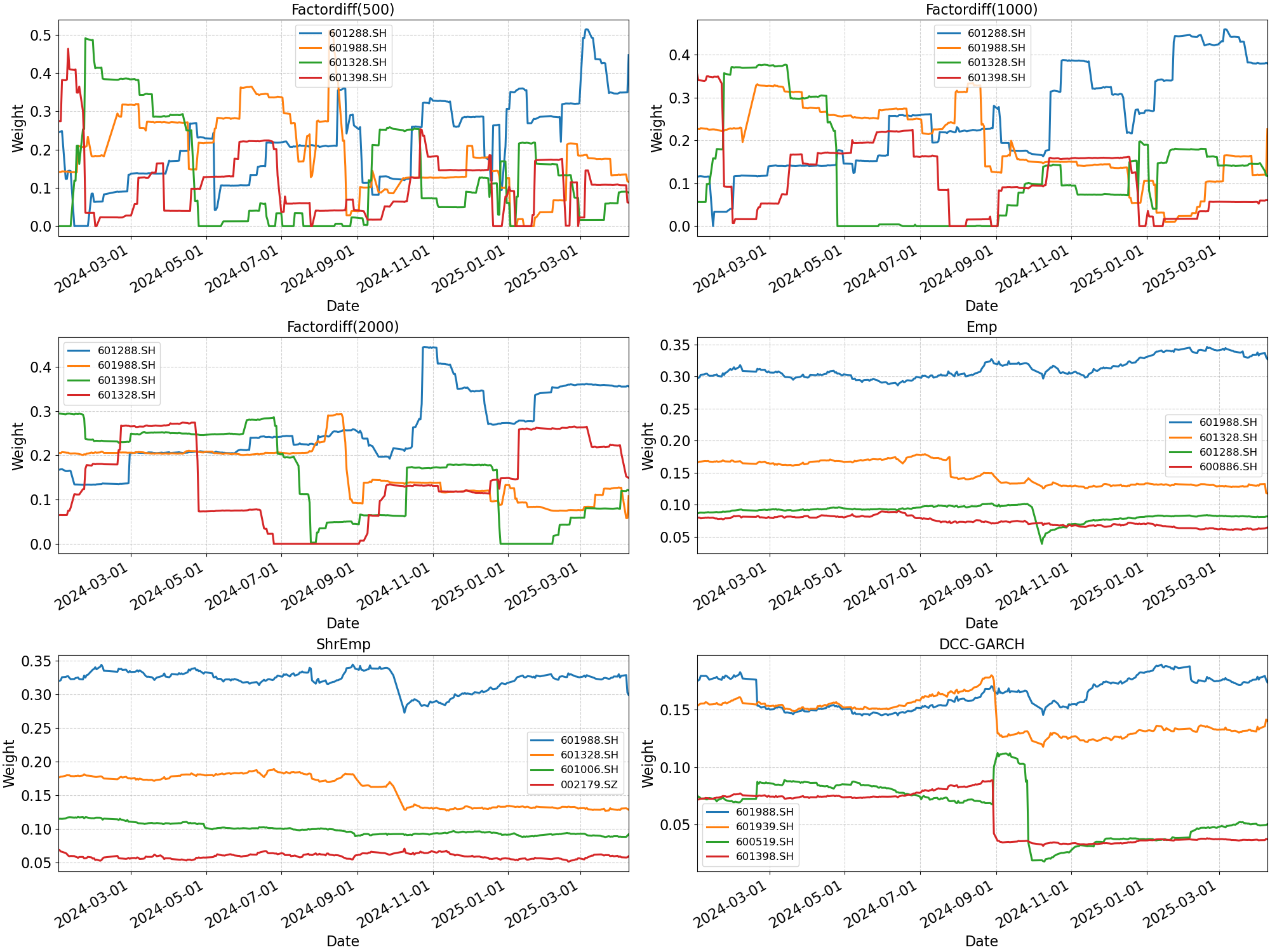} 
  \caption{Portfolio weights over time for the top 4 stocks in the optimal portfolio of \eqref{eq:mean_CVaR_tc} with stock return scenarios from Factordiff (500, 1000, and 2000 samples), Emp, ShrEmp, and DCC-GARCH.}
  \label{fig:weight_cvar_l1}
\end{figure}

The above findings again underscore the importance of accounting for 
transaction fees in the construction and evaluation of portfolio  strategies. Analogous to the mean-variance case, we incorporate transaction fees directly into the optimization objective:
\begin{equation}\label{eq:mean_CVaR_tc_mid}
\begin{aligned}
\max_{\bm{\omega}, \bm{b}, \bm{s}} \quad & \bm{\omega}^\top 
\boldsymbol{\mu_{t+1}} - \frac{\Gamma}{2} \mathrm{CVaR}_\beta(-\bm{\omega}^\top R_{t+1}) - \left(0.00075\, 
\bm{b}^\top \mathbf{1} + 0.00125\, \bm{s}^\top \mathbf{1} 
\right) \\
\text{s.t.} \quad & \bm{\omega}^\top \mathbf{1} = 1, 
 0\leq \omega_{i} \leq 1, \\
 & \bm{b} \geq 0, \bm{s} \geq 0,\\
&\omega_{i} - \omega^{d}_{i} = b_{i} - s_{i}, \quad \forall i,
\end{aligned}
\end{equation}
where the definition of $\bm{b}$, $\bm{s}$ and $\omega^{d}_{i}$ follow that of~\eqref{eq:mean_variance_tc}. To solve \eqref{eq:mean_CVaR_tc_mid}, we use samples of $R_{t+1}$ and rewrite it as:
\begin{equation}\label{eq:mean_CVaR_tc}
\begin{aligned}
\max_{\bm{\omega},\, \bm{b},\, \bm{s},\,\alpha, \bm{z}} \quad 
  & \bm{\omega}^\top \boldsymbol{\mu_{t+1}} 
    - \frac{\Gamma}{2}\left( \alpha + \frac{1}{M(1 - \beta)} \sum_{j=1}^{M} z_j \right)
    - \left(0.00075\, \bm{b}^\top \mathbf{1} 
    + 0.00125\, \bm{s}^\top \mathbf{1}\right) \\
\text{s.t.} \quad & \bm{\omega}^\top \mathbf{1} = 1, 
 0\leq \omega_{i} \leq 1, \\
 & \bm{b} \geq 0, \bm{s} \geq 0,\\
&\omega_{i} - \omega^{d}_{i} = b_{i} - s_{i}, \quad \forall i\\
& z_j + \bm{\omega}^\top r_{t+1,j} + \alpha \geq 0, \quad \forall j ,\\
& z_j \geq 0, \quad \forall j,
\end{aligned}
\end{equation}
the definition of $M$ and $r_{t+1,j}$ follow that of~\eqref{eq:mean_CVaR_lq}. The results are presented in Table~\ref{cvar-l1}, where we can see that Factordiff consistently outperforms the other strategies across all risk-adjusted performance metrics. In particular, Factordiff~(2000) achieves the highest mean return (0.109\%), Sharpe ratio (0.096), Sortino ratio (0.140), Calmar ratio (0.012), and RtC (0.042), substantially outperforming Emp, ShrEmp, and DCC-GARCH.  The performance of Factordiff is stable across all 
three sample sizes, with all of them outperforming the benchmark 
methods. As shown in Figure~\ref{fig:weight_cvar_l1}, incorporating transaction fees into the objective substantially smooths the portfolio weights of both Factordiff and DCC-GARCH relative to the fee-free case (Figure~\ref{fig:weight_cvar}). Among the Factordiff variants, Factordiff~(1000) and Factordiff~(2000) exhibit noticeably lower turnover than Factordiff~(500), indicating that the sample-based estimates of the mean return and CVaR stabilize only when the number of generated samples is sufficiently large. This is 
especially critical for CVaR, which depends on the tail of the 
loss distribution and therefore requires more samples 
than the mean does to achieve a reliable estimate. The above results confirm that the superiority of 
our conditional diffusion model extends beyond mean-variance 
optimization to mean-CVaR portfolio selection, and that incorporating 
transaction costs into the optimization objective is essential for 
realizing its full potential.

\begin{table}[H]
  \caption{Performance of the EW portfolio and the optimal portfolio 
  of \eqref{eq:mean_CVaR_lq} (with $\Gamma=1$) with Factordiff 
  (500, 1000, and 2000 samples), Emp, ShrEmp, and DCC-GARCH estimates 
  of stock return distribution. Transaction fees are deducted.}
  \label{cvar_with_cost}
  \centering
  \small
  \begin{tabular}{l|cccccc}
    \toprule
    Method & Mean (\%) & Std (\%) & Sharpe ratio & Sortino ratio 
           & Calmar ratio & RtC \\
    \midrule
    EW                 & 0.043              & 1.350              & 0.032              & 0.048              & 0.003            & 0.015              \\
    Factordiff (500)   & 0.022           & 1.081           & 0.020           
                       & 0.029           & 0.002                  
                       & 0.009           \\
    Factordiff (1000)  & 0.041           & 1.116           & 0.037           
                       & 0.054           & 0.004                  
                       & 0.016           \\
    Factordiff (2000)  & 0.046           & 1.114           & 0.042           
                       & 0.060           & 0.005                   
                       & 0.018           \\
    Emp                & 0.080           & $\bm{0.974}$   & $\bm{0.082}$   
                       & 0.121           & 0.009             
                       & 0.036           \\
    ShrEmp             & 0.076           & 0.986           & 0.077           
                       & 0.112           & 0.008                 
                       & 0.034           \\
    DCC-GARCH          & $\bm{0.084}$   & 1.034          & 0.081          
                       & $\bm{0.128}$   & $\bm{0.010}$          
                       & $\bm{0.040}$   \\
    \bottomrule
  \end{tabular}
\end{table}

In summary, our experiments on the Chinese A-share market demonstrate that the proposed factor-based conditional diffusion model consistently outperforms standard empirical, shrinkage, and DCC-GARCH estimators, as well as the EW baseline. Importantly, this superiority holds true for both mean-variance and mean-CVaR frameworks, under a realistic setting where transaction costs are explicitly incorporated into the optimization objective and the subsequent performance evaluation.

\begin{table}[H]
  \caption{Performance of the EW portfolio and the optimal portfolio 
  of \eqref{eq:mean_CVaR_tc} (with $\Gamma=1$) with Factordiff 
  (500, 1000, and 2000 samples), Emp, ShrEmp, and DCC-GARCH estimates 
  of stock return distribution. Transaction fees are incorporated into the 
  optimization objective and deducted.}
  \label{cvar-l1}
  \centering
  \small
  \begin{tabular}{l|cccccc}
    \toprule
    Method & Mean (\%) & Std (\%) & Sharpe ratio & Sortino ratio 
           & Calmar ratio  & RtC \\
    \midrule
    EW                 & 0.043              & 1.350              & 0.032              & 0.048              & 0.003            & 0.015              \\
    Factordiff (500)   & 0.102           & 1.123           & 0.091           
                       & 0.135           & 0.011                   
                       & 0.041           \\
    Factordiff (1000)  & 0.099           & 1.135           & 0.088           
                       & 0.129           & 0.011                  
                       & 0.039           \\
    Factordiff (2000)  & $\bm{0.109}$   & 1.141           & $\bm{0.096}$   
                       & $\bm{0.140}$   & $\bm{0.012}$          
                       & $\bm{0.042}$   \\
    Emp                & 0.080           & $\bm{0.984}$   & 0.081           
                       & 0.119           & 0.009                   
                       & 0.035           \\
    ShrEmp             & 0.080           & 0.996           & 0.080           
                       & 0.117           & 0.008                  
                       & 0.035           \\
    DCC-GARCH          & 0.078           & 1.008          & 0.078          
                       & 0.119         & 0.009          
                       & 0.037          \\
    \bottomrule
  \end{tabular}
\end{table}

\section{Error Analysis}\label{sec:sensitivity}

This section establishes theoretical error guarantees for our approach. After presenting preliminaries in Section~\ref{sec:setup}, Section~\ref{sec:sensitvity} derives explicit suboptimality bounds for both contextual mean-variance and mean-CVaR problems. Specifically, we bound the gap between the true optimal objective and the true objective yielded by the diffusion-based solution, demonstrating that it scales linearly with the 2-Wasserstein distance (\(\mathcal{W}_{2}\)) between the two distributions. Lastly, Section~\ref{sec:W-error} provides a \(\mathcal{W}_{2}\) error bound for the conditional diffusion model, directly connecting this distributional gap to the model's training loss.



\subsection{Setup and preliminary results}\label{sec:setup}

Fix time $t$. Let $P_x$ denote the true (unknown) conditional distribution of $R_{t+1}\in\mathbb{R}^{D}$ given the factor matrix $X_t =x$, and let $Q_x$ denote the distribution learned by our conditional diffusion model. We write $R\sim P_x$ and $\tilde{R}\sim Q_x$ to denote return random vectors drawn from the respective distributions. Define the moments:
\begin{align*}
\boldsymbol{\mu}_{P,x} = \mathbb{E}_{P_x}[R],\quad &\boldsymbol{\Sigma}_{P,x} = \mathrm{Cov}_{P_x}(R),\\
\boldsymbol{\mu}_{Q,x} = \mathbb{E}_{Q_x}[\tilde{R}],\quad &\boldsymbol{\Sigma}_{Q,x} = \mathrm{Cov}_{Q_x}(\tilde{R}).
\end{align*}
The 2-Wasserstein distance between $P_x$ and $Q_x$ is 
\[
\mathcal{W}_2(P_x,Q_x) = \left(\inf_{\pi\in\Pi(P_x,Q_x)}\mathbb{E}_{(R,\tilde{R})\sim\pi}\left[\|R-\tilde{R}\|_2^2\right]\right)^{1/2},
\]
where $\Pi(P_x,Q_x)$ denotes the set of all couplings (joint distributions) with marginals $P_x$ and $Q_x$ \citep{villani2009optimal}. An \emph{optimal coupling} $(R^*,\tilde{R}^*)$ is a pair satisfying $R^*\sim P_x$, $\tilde{R}^*\sim Q_x$, and $\mathbb{E}[\|R^*-\tilde{R}^*\|_2^2]=\mathcal{W}_2(P_x,Q_x)^2$. 
We adopt the following notations: $\|\mathbf{v}\|_\infty = \max_k |v_k|$ for a vector $\mathbf{v}$, and $\|A\|_{\max}=\max_{i,j}|A_{ij}|$ for a matrix $A$. To proceed with our analysis, we impose the following assumption.

\begin{assumption}\label{assump:moment}
For all $x$, the distributions $P_x$ and $Q_x$ are in distribution family $\mathcal{P}$ with bounded second moments, i.e., $M \coloneqq \sup_{p \in \mathcal{P}} \mathbb{E}_{\textbf{R} \sim p} [\lVert \textbf{R} \rVert_2^2] < \infty$. 
\end{assumption}

The following result bounds the differences in moments under $P_x$ and $Q_x$ in terms of $\mathcal{W}_2(P_x,Q_x)$. These are standard results that can be inferred from the literature (e.g., \cite{gelbrich1990formula, villani2009optimal}).

\begin{lemma}\label{lemma:meanvar}
For any $x$, we have
\begin{itemize}
\item [(a)] $\|\boldsymbol{\mu}_{P,x} - \boldsymbol{\mu}_{Q,x}\|_{\infty} \le \mathcal{W}_2(P_x, Q_x).$

\item [(b)] $\|\boldsymbol{\Sigma}_{P,x} - \boldsymbol{\Sigma}_{Q,x}\|_{\max} \le 4\sqrt{M}\cdot\mathcal{W}_2(P_x, Q_x),$ provided that Assumption~\ref{assump:moment} holds.

\end{itemize}
\end{lemma}

\subsection{Sensitivity analysis of mean-variance and mean-CVaR portfolios}\label{sec:sensitvity}

We first consider the mean-variance portfolio optimization problem with transaction costs~\eqref{eq:mean_variance_tc} and conduct the sensitivity analysis. Under the true conditional distribution $P_x$ and the learned conditional distribution $Q_x$, the objective functions are denoted by
\begin{align*}
G_{P_x}^{\mathrm{MV}}(\bm{\omega}, \bm{b}, \bm{s}) &= \bm{\omega}^\top \boldsymbol{\mu}_{P,x} - \tfrac{\gamma}{2}\,\bm{\omega}^\top\boldsymbol{\Sigma}_{P,x}\bm{\omega} - \big(0.00075\,\bm{b}^\top\mathbf{1}+0.00125\,\bm{s}^\top\mathbf{1}\big),\\
G_{Q_x}^{\mathrm{MV}}(\bm{\omega}, \bm{b}, \bm{s}) &= \bm{\omega}^\top \boldsymbol{\mu}_{Q,x} - \tfrac{\gamma}{2}\,\bm{\omega}^\top\boldsymbol{\Sigma}_{Q,x}\bm{\omega} - \big(0.00075\,\bm{b}^\top\mathbf{1}+0.00125\,\bm{s}^\top\mathbf{1}\big).
\end{align*}
Let $\mathcal{F}$ denote the feasible set defined by the constraints in~\eqref{eq:mean_variance_tc}, let $(\bm{\omega}_{P,x}^*,\bm{b}_{P,x}^*,\bm{s}_{P,x}^*)$ be the optimal solutions associated with the objective $G_{P_x}^{\mathrm{MV}}(\bm{\omega}, \bm{b}, \bm{s})$ when the true conditional moments $(\boldsymbol{\mu}_{P,x},\boldsymbol{\Sigma}_{P,x})$ are used, and let $(\bm{\omega}_{Q,x}^*,\bm{b}_{Q,x}^*,\bm{s}_{Q,x}^*)$ be the optimal solutions associated with $G_{Q_x}^{\mathrm{MV}}(\bm{\omega}, \bm{b}, \bm{s})$ when the learned conditional moments $(\boldsymbol{\mu}_{Q,x},\boldsymbol{\Sigma}_{Q,x})$ are used. With slight abuse of notations, we use $X$ to denote the distribution of the factor matrix at each time $t$, and $x$ is a sample from $X$. The following result provides an explicit error bound for the suboptimality of the learned optimal solution $(\bm{\omega}_{Q,x}^*,\bm{b}_{Q,x}^*,\bm{s}_{Q,x}^*)$ under the true model $P_X$.


\begin{theorem}[Suboptimality Gap for the Mean-Variance Problem]\label{thm:mv_sensitivity}
Suppose Assumption~\ref{assump:moment} holds. We have
\begin{align*}
\mathbb{E}_X \left[ \big|G_{P_X}^{\mathrm{MV}}(\bm{\omega}_{P_X}^*,\bm{b}_{P_X}^*,\bm{s}_{P_X}^*) - G_{P_X}^{\mathrm{MV}}(\bm{\omega}_{Q_X}^*,\bm{b}_{Q_X}^*,\bm{s}_{Q_X}^*)\big| \right] \le 2\big(1+2\gamma\sqrt{M}\big)\,\mathbb{E}_X \left[\mathcal{W}_2(P_X,Q_X)\right].
\end{align*}
\end{theorem}

We next consider the mean-CVaR portfolio optimization problem with transaction costs~\eqref{eq:mean_CVaR_tc_mid}, and analyze the error when we use a learned distribution for optimization. Denote by
\begin{align*}
G_{P_x}^{\mathrm{CVaR}}(\bm{\omega}, \bm{b}, \bm{s}) &= \bm{\omega}^\top \boldsymbol{\mu}_{P,x} - \tfrac{\Gamma}{2}\,\mathrm{CVaR}_\beta(-\bm{\omega}^\top R) - \big(0.00075\,\bm{b}^\top\mathbf{1}+0.00125\,\bm{s}^\top\mathbf{1}\big),\\
G_{Q_x}^{\mathrm{CVaR}}(\bm{\omega}, \bm{b}, \bm{s}) &= \bm{\omega}^\top \boldsymbol{\mu}_{Q,x} - \tfrac{\Gamma}{2}\,\mathrm{CVaR}_\beta(-\bm{\omega}^\top \tilde{R}) - \big(0.00075\,\bm{b}^\top\mathbf{1}+0.00125\,\bm{s}^\top\mathbf{1}\big),
\end{align*}
where $R\sim P_x$ and $\tilde{R}\sim Q_x$. We need the following bound on the CVaR difference, the proof of which is given in Appendix~\ref{sec:proof_thm1_thm2}.

\begin{lemma}[Bound on the CVaR Difference]\label{prop:cvar_bound}
For any portfolio weight vector $\bm{\omega}$ satisfying $\|\bm{\omega}\|_1=1$,
\[
\big|\mathrm{CVaR}_\beta(-\bm{\omega}^\top R) - \mathrm{CVaR}_\beta(-\bm{\omega}^\top \tilde{R})\big| \le \frac{1}{\sqrt{1-\beta}}\,\mathcal{W}_2(P_x,Q_x).
\]
\end{lemma}

Let $(\bar{\bm{\omega}}_{P,x}^*,\bar{\bm{b}}_{P,x}^*,\bar{\bm{s}}_{P,x}^*)$ and $(\bar{\bm{\omega}}_{Q,x}^*,\bar{\bm{b}}_{Q,x}^*,\bar{\bm{s}}_{Q,x}^*)$ denote the optimal solutions of~\eqref{eq:mean_CVaR_tc_mid} under $P_x$ and $Q_x$, respectively. Then we have the following result.

\begin{theorem}[Suboptimality Gap for the Mean-CVaR Problem]\label{thm:cvar_sensitivity} The following inequality holds:
\begin{align*}
\mathbb{E}_X \left[ \big|G_{P_X}^{\mathrm{CVaR}}(\bar{\bm{\omega}}_{P,X}^*,\bar{\bm{b}}_{P,X}^*,\bar{\bm{s}}_{P,X}^*) - G_{P_X}^{\mathrm{CVaR}}(\bar{\bm{\omega}}_{Q,X}^*,\bar{\bm{b}}_{Q,X}^*,\bar{\bm{s}}_{Q,X}^*)\big| \right]\le 2\left(1+\frac{\Gamma}{2\sqrt{1-\beta}}\right) \mathbb{E}_X \left[ \mathcal{W}_2(P_X,Q_X)\right].
\end{align*}
\end{theorem}

Theorems~\ref{thm:mv_sensitivity} and~\ref{thm:cvar_sensitivity} (proofs are deferred to Appendix~\ref{sec:proof_thm1_thm2}) show that the gap between the objective values under the true solution and the learned solution vanishes linearly as $\mathbb{E}_X \left[\mathcal{W}_2(P_X,Q_X)\right]\to 0$. This provides a theoretical justification for using the distribution generated by our conditional diffusion model as a proxy for the true conditional return distribution in portfolio optimization: as the generative model improves its distributional approximation, the resulting portfolio decisions approach optimality. 
A remaining, non-trivial question is whether one can establish an error bound for  $\mathbb{E}_X \left[\mathcal{W}_2(P_X,Q_X)\right]$ that directly links this $\mathcal{W}_2$ distance to the training loss of the conditional diffusion model in \eqref{eq:loss}. We address this question in the next section.



\subsection{Wasserstein error analysis of the conditional diffusion model}\label{sec:W-error}






We now establish the desired $\WassD$ error bound between $P_X$ and $Q_X$. To this end, we first state an assumption which imposes regularity conditions on the true return distribution, the training loss of the conditional diffusion model, as well as the corresponding learned return distribution. Recall the statistical factor model in \eqref{eq:model}.  


\begin{assumption}\label{assump:diffusion_model}
\begin{enumerate}
    \item[(1)] The distribution of the random shock $u_{t+1}$ in \eqref{eq:model} admits a twice continuously differentiable positive density $p_u (\cdot)$ on $\mathbb{R}^D$, and $\nabla \log p_u(\cdot)$ is $L_u$-Lipschitz continuous.

    \item[(2)] The true return $R_{t + 1}$ and generated return $\tilde{R}$ have finite third absolute moments, i.e., there exists a positive constant $M_3$ such that  
    \begin{align}
       & \mathbb{E}_{x \sim X_t}\big[\mE_{R_{t + 1}\sim P_x}[\|R_{t + 1}\|^3]\big] = \mathbb{E}_{x \sim X_t} \big[\mE[\|f(x) + u_{t + 1}\|^3]\big] \le M_3, \label{eq:assum_tar_moment_3} \\
        & \mE_{x\sim X_t}\big[\mE_{\tilde{R}\sim Q_x}||\tilde{R}||^3\big] \le {M}_3. \label{eq:assum_gen_moment_3}
    \end{align}
   \item [(3)] There exists a small positive constant $\varepsilon_{\text{noise}}$ such that the optimized training loss corresponding to the trained neural network $\bepsilon_{\hat\theta^*}$ satisfies 
    \begin{align}\label{eq:assum_DSM_error}
        L(\hat\theta^*) = \frac{1}{N}\sum_{n=1}^N  \mathbb{E}_{(X_{t},R_{t+1}),\epsilon}\left[\|\epsilon-\bepsilon_{\hat\theta^*}\left(R^{(n)}_{t+1},n;X_{t}\right)\|^2 \right]\le \varepsilon_{\text{noise}}^2.
    \end{align}

\end{enumerate}

\end{assumption}

We briefly discuss the implications of Assumption~\ref{assump:diffusion_model}. Part (1) imposes standard regularity conditions on the random shock, which are satisfied by widely used distributions in the quantitative finance literature, including multivariate Gaussian, Gaussian mixtures, and Student-\(t\) distributions. 
Part (2) assumes a finite third absolute moment for the daily return distribution (this assumption can be weakened to a \((2+\delta)\)-th moment for small $\delta>0$; see the discussion at the end of this section). Moreover, 
it is natural to expect the generated return distribution to inherit this property under a well-behaved, trained network in the conditional diffusion model. We provide formal sufficient conditions ensuring that \eqref{eq:assum_gen_moment_3} holds in Appendix~\ref{sec:gen_dist_finite_moment}. 
Finally, Part (3) assumes that the conditional diffusion model is sufficiently optimized such that the training loss remains small. This assumption is standard for establishing theoretical error bounds and sampling guarantees in the diffusion literature \citep{chen2023sampling, bentonnearly, chen2023improved, gao2026data}.


With Assumption \ref{assump:diffusion_model}, we are now ready to derive a $\WassD$ error bound between $P_X$ and $Q_X$. The proof of Theorem \ref{thm:W2_error_bound} is given in Appendix \ref{sec:proof_W2_bound}. Recall that the return data has a dimension of \(D\) and $(\eta_i)_{i=1}^N$ is the variance/noise schedule in the DDPM model. 

\begin{theorem}\label{thm:W2_error_bound}
    Under Assumption \ref{assump:diffusion_model}, the following error bound holds: 
    \begin{align}
        &\mE_{X}\big[\WassD\big(P_X, Q_X \big) \big]\nonumber\\
        &\le C_{\WassD}\cdot {M_3}^{1/3}\cdot \bigg(\underbrace{e^{-\frac{N}{2}(\eta_1 + \eta_N)} \cdot \big(M_3^{2/3} + D\big)}_{(i)} + \underbrace{\frac{1}{N} \big(M_3^{2/3} + D + DL_u\big)}_{(ii)} + \underbrace{\frac{\varepsilon_{\text{noise}}^2}{1 - e^{- \eta_1}}}_{(iii)}\bigg)^{1/12}, \label{eq:W2_bound}
    \end{align}
    for some positive constant $C_{\WassD}$ independent of $N$, $D$, $L_u, M_3$, and $\varepsilon_{\text{noise}}$.
\end{theorem}

Theorem \ref{thm:W2_error_bound} characterizes the performance guarantees of our conditional diffusion model using the 2-Wasserstein ($\WassD$) distance. 
The bound \eqref{eq:W2_bound} cleanly separates three distinct sources of error: 
\begin{itemize}
\item Firstly, the term (i) represents the initialization mismatch error between $R^{(N)}_{t + 1}$ in \eqref{eq:forward} and $\widetilde{R}^{(N)}_{t + 1}$ in \eqref{eq:reverse}. 
This term decays exponentially and becomes negligible when \(N(\eta_1 + \eta_N)\) is sufficiently large. 

\item Next, term (ii) accounts for the discretization error incurred by approximating the underlying continuous-time denoising dynamics with the discrete reverse process \eqref{eq:reverse}. Utilizing the continuous-time framework of diffusion models in \cite{songscore} facilitates this error analysis. Since the discretization step size scales as \(1/N\), this error term vanishes as the number of total steps \(N\) becomes large.

\item Finally, term (iii) reflects the impact of the trained neural network's inaccuracy during the denoising/sampling process. This term scales directly with the optimized training loss, and remains small when the diffusion model is well-trained. The pre-factor \((1-e^{-\eta_1})^{-1}\) arises from translating the discrete-time noise prediction loss in \eqref{eq:loss} into the continuous-time score-matching loss in \cite{songscore}.



\end{itemize}

We note that the bound \eqref{eq:W2_bound} can be further tightened under stronger assumptions. Specifically, if we assume the true return $R_{t + 1}$ and generated return $\tilde{R}$ have finite $q$-th absolute moments for some \(q>2\) ($q$ is not necessarily an integer),  our proof still holds and the exponent in the bound \eqref{eq:W2_bound} changes from \(\frac{1}{12}\) to \(\frac{q-2}{4q}\). Consequently, the error bound sharpens for $q>3$ but deteriorates as \(q\) approaches 2. Furthermore, although the factor dimension \(K\) does not appear explicitly in Theorem~\ref{thm:W2_error_bound}, it enters the bound implicitly through the constants \(M_{3}\) and \(\varepsilon _{\text{noise}}^{2}\) due to their dependence on the factor matrix \(X_{t}\). Together, Theorem~\ref{thm:W2_error_bound}, along with Theorems~\ref{thm:mv_sensitivity} and~\ref{thm:cvar_sensitivity},
provides an end-to-end guarantee for our proposed approach to contextual portfolio optimization: as the conditional diffusion model minimizes its training loss, our approach yields portfolio decisions that converge toward optimality.

\section{Conclusions}\label{sec:conclusion}

This paper developed a factor-based conditional diffusion model that generates the full cross-sectional distribution of next-day stock returns conditioned on asset-specific observable factors. Experiments on the Chinese A-share market revealed that incorporating transaction costs directly into the optimization objective is essential, and the empirical results demonstrate that our diffusion-based approach  consistently outperform various benchmarks in both mean-variance and mean-CVaR portfolios. The accompanying error analysis provides theoretical support of our approach.


This work opens the gate to several directions of future research. First, it will be important to empirically evaluate our generative approach and test its robustness across diverse asset classes and markets. Second, integrating richer conditioning information, such as fundamental data or market regimes, could potentially enhance model performance. Finally, it would be interesting to extend the application of conditional generative diffusion models to broader classes of risk-sensitive contextual stochastic optimization problems.

    \newpage
	\bibliographystyle{plainnat}
	\bibliography{ref}

\newpage 

\appendix

\section{Appendix}

\subsection{Performance Evaluation Metrics}\label{app:experiment_setting}

We use $r_{p, t}$ to denote the return of the constructed portfolio on 
day $t$ over the test period of $T$ trading days. Assume the risk free rate is 0, the evaluation metrics are defined as follows.

\begin{itemize}
\item  \textbf{Mean return.} $\mu_p=\frac{1}{T}\sum_{t=1}^{T}r_{p,t}$.
  \item \textbf{Standard deviation.}
   $\sigma_p = \sqrt{\frac{1}{T-1}\sum_{t=1}^{T}
        (r_{p,t}-\mu_p)^{2}}$.
  \item \textbf{Sharpe ratio.}
    The Sharpe ratio evaluates the 
    excess return earned per unit of total risk:
    \begin{equation*}
      \text{Sharpe Ratio} = \frac{\mu_p}{\sigma_p}.
    \end{equation*}

  \item \textbf{Sortino ratio.}
    Unlike the Sharpe ratio, which penalizes upside and downside 
    deviations equally, the Sortino ratio 
    isolates downside risk by 
    replacing $\sigma_p$ with the downside semi-deviation:
    \begin{equation*}
      \text{Sortino Ratio} = \frac{\mu_p }
        {\sigma_{\mathrm{down}}},
      \qquad
      \sigma_{\mathrm{down}}
        = \sqrt{\frac{1}{T}\sum_{t=1}^{T}\min(r_{p,t},0)^2}.
    \end{equation*}

  \item \textbf{Calmar ratio.}
    The Calmar ratio measures the return relative to the 
    maximum drawdown (MDD), which captures the largest peak-to-trough 
    decline of the cumulative portfolio value 
    $V_t=\prod_{k=1}^{t}(1+r_{p,k})$:
    \begin{equation*}
    \text{MDD} = \max_{t \in [T]}
        \frac{\max_{s \leq t} V_s - V_t}{\max_{s \leq t} V_s },
      \qquad
      \text{Calmar Ratio} = \frac{\mu_p}{\text{MDD}},
    \end{equation*}

  \item \textbf{Return-to-CVaR (RtC).}
The RtC ratio is defined as
\begin{equation*}
\text{RtC}
=
\frac{\mu_p}{\mathrm{CVaR}_\beta(-R_p)},
\end{equation*}
where $R_p$ denotes the empirical distribution of the portfolio return samples, and $\beta = 0.95$.
\end{itemize}

\subsection{Robustness Checks}\label{sec:robust}

To verify that our main findings are not sensitive to the choice of risk aversion parameters, we repeat the portfolio optimization experiments under alternative parameter values. 

Specifically, we set $\gamma=50$ for the mean-variance formulation~\eqref{eq:mean_variance_tc} and $\Gamma=0.5$ for the mean-CVaR formulation~\eqref{eq:mean_CVaR_tc}, both with transaction costs incorporated into the optimization objective. The results are reported in Tables~\ref{mv-l1_50} and~\ref{cvar-l1_05}, respectively. Under $\gamma=50$, Factordiff continues to dominate all benchmark methods across every risk-adjusted metric, with Factordiff~(500) achieving the highest Sharpe ratio (0.102), Sortino ratio (0.150), Calmar ratio (0.011), and RtC (0.045). Similarly, under $\Gamma=0.5$, all three Factordiff variants substantially outperform EW, Emp, ShrEmp, and DCC-GARCH. These results confirm that the superior performance of our factor-based conditional diffusion model is robust to the choice of risk aversion parameters in both the mean-variance and mean-CVaR frameworks.

\begin{table}
  \caption{Performance of the EW portfolio and the optimal portfolio 
  of \eqref{eq:mean_variance_tc} (with $\gamma=50$) with Factordiff 
  (500, 1000, and 2000 samples), Emp, ShrEmp, and DCC-GARCH estimates 
  of stock return moments. Transaction fees are incorporated into the 
  optimization objective and deducted.}
  \label{mv-l1_50}
  \centering
  \small
  \begin{tabular}{l|cccccc}
    \toprule
    Method & Mean (\%) & Std (\%) & Sharpe ratio& Sortino ratio& Calmar ratio& RtC \\
    \midrule
    EW                 & 0.043              & 1.350              & 0.032              & 0.048              & 0.003            & 0.015              \\
    Factordiff (500)   & $\bm{0.122}$           & 1.203           & $\bm{0.102}$           & $\bm{0.150}$          & $\bm{0.011}$   & $\bm{0.045}$   \\
    Factordiff (1000)  & 0.114           & 1.194           & 0.096           & 0.140           & 0.010           & 0.042           \\
    Factordiff (2000)  & 0.115   & 1.194           & 0.096   & 0.141   & 0.010          & 0.042           \\
    Emp                & 0.070           & $\bm{0.953}$   & 0.074          & 0.109           & 0.008           & 0.033           \\
    ShrEmp             & 0.078           & 0.978          & 0.079           & 0.115          & 0.008           & 0.034           \\
    DCC-GARCH          & 0.079           & 0.983           & 0.081          & 0.130           & 0.007          & 0.040          \\
    \bottomrule
  \end{tabular}
\end{table}

\begin{table}[H]
  \caption{Performance of the EW portfolio and the optimal portfolio 
  of \eqref{eq:mean_CVaR_tc} (with $\Gamma=0.5$) with Factordiff 
  (500, 1000, and 2000 samples), Emp, ShrEmp, and DCC-GARCH estimates 
  of stock return moments. Transaction fees are incorporated into the 
  optimization objective and deducted.}
  \label{cvar-l1_05}
  \centering
  \small
  \begin{tabular}{l|cccccc}
    \toprule
    Method & Mean (\%) & Std (\%) & Sharpe ratio & Sortino ratio 
           & Calmar ratio  & RtC \\
    \midrule
    EW                 & 0.043              & 1.350              & 0.032              & 0.048              & 0.003            & 0.015              \\
    Factordiff (500)   & 0.125           & 1.170          & 0.107           
                       & 0.161           & $\bm{0.014}$                   
                       & $\bm{0.049}$           \\
    Factordiff (1000)  & 0.123           & 1.200           & 0.103           
                       & 0.152           & 0.011                  
                       & 0.046           \\
    Factordiff (2000)  & $\bm{0.131}$   & 1.202           & $\bm{0.109}$   
                       & $\bm{0.161}$   & 0.012          
                       & 0.048   \\
    Emp                & 0.076           & 0.976   & 0.078           
                       & 0.114           & 0.008                   
                       & 0.034          \\
    ShrEmp             & 0.078           & 1.001           & 0.078           
                       & 0.113           & 0.008                  
                       & 0.034           \\
    DCC-GARCH          & 0.072          & $\bm{0.967}$          & 0.075         
                       & 0.115         & 0.008          
                       & 0.036         \\
    \bottomrule
  \end{tabular}
\end{table}




\subsection{Proofs of Results in Section~\ref{sec:sensitvity}}\label{sec:proof_thm1_thm2}

In this section, we provide the proofs of Theorem \ref{thm:mv_sensitivity}, Lemma~\ref{prop:cvar_bound}, and Theorem \ref{thm:cvar_sensitivity}.

\proof{Proof of Theorem \ref{thm:mv_sensitivity}.}
For any $(\bm{\omega},\bm{b},\bm{s})\in\mathcal{F}$, the constraints $\bm{\omega}^\top\mathbf{1}=1$ and $\omega_i\ge 0$ imply $\|\bm{\omega}\|_1=1$. By the triangle inequality and the definition of $\|\cdot\|_{\max}$,
\begin{align}
\big|G_{P_x}^{\mathrm{MV}}(\bm{\omega},\bm{b},\bm{s}) - G_{Q_x}^{\mathrm{MV}}(\bm{\omega},\bm{b},\bm{s})\big| &= \big|\bm{\omega}^\top(\boldsymbol{\mu}_{P,x}-\boldsymbol{\mu}_{Q,x}) - \tfrac{\gamma}{2}\,\bm{\omega}^\top(\boldsymbol{\Sigma}_{P,x}-\boldsymbol{\Sigma}_{Q,x})\bm{\omega}\big|\nonumber\\
&\le \|\bm{\omega}\|_1\|\boldsymbol{\mu}_{P,x}-\boldsymbol{\mu}_{Q,x}\|_\infty + \tfrac{\gamma}{2}\|\bm{\omega}\|_1^2\|\boldsymbol{\Sigma}_{P,x}-\boldsymbol{\Sigma}_{Q,x}\|_{\max}\nonumber\\
&= \|\boldsymbol{\mu}_{P,x}-\boldsymbol{\mu}_{Q,x}\|_\infty + \tfrac{\gamma}{2}\|\boldsymbol{\Sigma}_{P,x}-\boldsymbol{\Sigma}_{Q,x}\|_{\max}.\label{eq:mv_pointwise}
\end{align}
Applying Lemma~\ref{lemma:meanvar}, we obtain that for any feasible $(\bm{\omega},\bm{b},\bm{s})\in\mathcal{F}$,
\begin{equation}\label{eq:mv_uniform}
\big|G_{P_x}^{\mathrm{MV}}(\bm{\omega},\bm{b},\bm{s}) - G_{Q_x}^{\mathrm{MV}}(\bm{\omega},\bm{b},\bm{s})\big| \le (1+2\gamma\sqrt{M})\,\mathcal{W}_2(P_x,Q_x).
\end{equation}
We now bound the suboptimality and apply the following decomposition: 
\begin{align*}
&G_{P_x}^{\mathrm{MV}}(\bm{\omega}_{P,x}^*,\bm{b}_{P,x}^*,\bm{s}_{P,x}^*) - G_{P_x}^{\mathrm{MV}}(\bm{\omega}_{Q,x}^*,\bm{b}_{Q,x}^*,\bm{s}_{Q,x}^*) \\
&\quad= \underbrace{\big[G_{P_x}^{\mathrm{MV}}(\bm{\omega}_{P,x}^*,\bm{b}_{P,x}^*,\bm{s}_{P,x}^*)-G_{Q_x}^{\mathrm{MV}}(\bm{\omega}_{P,x}^*,\bm{b}_{P,x}^*,\bm{s}_{P,x}^*)\big]}_{\text{Term I}}\\
&\qquad+\underbrace{\big[G_{Q_x}^{\mathrm{MV}}(\bm{\omega}_{P,x}^*,\bm{b}_{P,x}^*,\bm{s}_{P,x}^*) - G_{Q_x}^{\mathrm{MV}}(\bm{\omega}_{Q,x}^*,\bm{b}_{Q,x}^*,\bm{s}_{Q,x}^*)\big]}_{\text{Term II}}\\
&\qquad+\underbrace{\big[G_{Q_x}^{\mathrm{MV}}(\bm{\omega}_{Q,x}^*,\bm{b}_{Q,x}^*,\bm{s}_{Q,x}^*) - G_{P_x}^{\mathrm{MV}}(\bm{\omega}_{Q,x}^*,\bm{b}_{Q,x}^*,\bm{s}_{Q,x}^*)\big]}_{\text{Term III}}.
\end{align*}
For Term~I, applying~\eqref{eq:mv_uniform} at the feasible point $(\bm{\omega}_{P,x}^*,\bm{b}_{P,x}^*,\bm{s}_{P,x}^*)$ yields
\[
\text{Term I} \le (1+2\gamma\sqrt{M})\,\mathcal{W}_2(P_x,Q_x).
\]
For Term~II, by the optimality of $(\bm{\omega}_{Q,x}^*,\bm{b}_{Q,x}^*,\bm{s}_{Q,x}^*)$ for $G_{Q_x}^{\mathrm{MV}}$ over $\mathcal{F}$, we have
\[
\text{Term II} = G_{Q_x}^{\mathrm{MV}}(\bm{\omega}_{P,x}^*,\bm{b}_{P,x}^*,\bm{s}_{P,x}^*) - G_{Q_x}^{\mathrm{MV}}(\bm{\omega}_{Q,x}^*,\bm{b}_{Q,x}^*,\bm{s}_{Q,x}^*) \le 0.
\]
For Term~III, applying~\eqref{eq:mv_uniform} at the feasible point $(\bm{\omega}_{Q,x}^*,\bm{b}_{Q,x}^*,\bm{s}_{Q,x}^*)$ yields
\[
\text{Term III} \le (1+2\gamma\sqrt{M})\,\mathcal{W}_2(P_x,Q_x).
\]
Combining the three terms, we obtain
\begin{equation}\label{eq:mvx}
0 \leq G_{P_x}^{\mathrm{MV}}(\bm{\omega}_{P,x}^*,\bm{b}_{P,x}^*,\bm{s}_{P,x}^*) - G_{P_x}^{\mathrm{MV}}(\bm{\omega}_{Q,x}^*,\bm{b}_{Q,x}^*,\bm{s}_{Q,x}^*) \le 2(1+2\gamma\sqrt{M})\,\mathcal{W}_2(P_x,Q_x).
\end{equation}
Since~\eqref{eq:mvx} holds for every $x$, taking expectations on both sides with respect to $X$ yields 
\begin{equation*}
\mathbb{E}_X \left[ \big|G_{P_X}^{\mathrm{MV}}(\bm{\omega}_{P_X}^*,\bm{b}_{P_X}^*,\bm{s}_{P_X}^*) - G_{P_X}^{\mathrm{MV}}(\bm{\omega}_{Q_X}^*,\bm{b}_{Q_X}^*,\bm{s}_{Q_X}^*)\big| \right] \le 2\big(1+2\gamma\sqrt{M}\big)\,\mathbb{E}_X \left[\mathcal{W}_2(P_X,Q_X)\right].
\end{equation*}
This completes the proof.
\endproof
\proof{Proof of Lemma~\ref{prop:cvar_bound}.}
Let the portfolio loss function be defined as \(f(R) = -\bm{\omega}^\top R\). For any two return vectors \(R_1, R_2\), by the Cauchy--Schwarz inequality, we have
\[
|f(R_1) - f(R_2)| = |\bm{\omega}^\top (R_2 - R_1)| \le \|\bm{\omega}\|_2 \|R_1 - R_2\|_2.
\]
Thus, the Lipschitz constant of the function \(f\) with respect to the \(\ell_2\)-norm is \(L = \|\bm{\omega}\|_2\).

By applying Corollary 11 of \citep{pichler2013evaluations} directly to the  \(P_x\) and \(Q_x\), we obtain the following bound:
\[
\big|\mathrm{CVaR}_\beta(-\bm{\omega}^\top R) - \mathrm{CVaR}_\beta(-\bm{\omega}^\top \tilde{R})\big| \le L \cdot \frac{1}{\sqrt{1-\beta}}\,\mathcal{W}_2(P_x, Q_x).
\]
Since \(\|\bm{\omega}\|_1 = 1\), we have \(L = \|\bm{\omega}\|_2 \le \|\bm{\omega}\|_1 = 1\). Substituting \(L \le 1\) into the inequality above yields the desired bound:
\[
\big|\mathrm{CVaR}_\beta(-\bm{\omega}^\top R) - \mathrm{CVaR}_\beta(-\bm{\omega}^\top \tilde{R})\big| \le \frac{1}{\sqrt{1-\beta}}\,\mathcal{W}_2(P_x, Q_x). 
\]
This completes the proof.
\endproof

\proof{Proof of Theorem \ref{thm:cvar_sensitivity}.}
For any feasible $(\bm{\omega},\bm{b},\bm{s})$ with $\|\bm{\omega}\|_1=1$, the triangle inequality gives
\begin{align}
\big|G_{P_x}^{\mathrm{CVaR}}(\bm{\omega},\bm{b},\bm{s})-G_{Q_x}^{\mathrm{CVaR}}(\bm{\omega},\bm{b},\bm{s})\big| &\le |\bm{\omega}^\top(\boldsymbol{\mu}_{P,x}-\boldsymbol{\mu}_{Q,x})|+\frac{\Gamma}{2}\big|\mathrm{CVaR}_\beta(-\bm{\omega}^\top R)-\mathrm{CVaR}_\beta(-\bm{\omega}^\top \tilde{R})\big|\nonumber\\
&\le \|\boldsymbol{\mu}_{P,x}-\boldsymbol{\mu}_{Q,x}\|_\infty + \frac{\Gamma}{2\sqrt{1-\beta}}\,\mathcal{W}_2(P_x,Q_x)\nonumber\\
&\le \left(1+\frac{\Gamma}{2\sqrt{1-\beta}}\right)\mathcal{W}_2(P_x,Q_x),\label{eq:cvar_pointwise}
\end{align}
where we used Lemma ~\ref{lemma:meanvar} and Proposition ~\ref{prop:cvar_bound}. 
The suboptimality is non-negative by the optimality of $(\bar{\bm{\omega}}_{P,x}^*,\bar{\bm{b}}_{P,x}^*,\bar{\bm{s}}_{P,x}^*)$ for $G_{P_x}^{\mathrm{CVaR}}$. We decompose it as follows:
\begin{align*}
&G_{P_x}^{\mathrm{CVaR}}(\bar{\bm{\omega}}_{P,x}^*,\bar{\bm{b}}_{P,x}^*,\bar{\bm{s}}_{P,x}^*) - G_{P_x}^{\mathrm{CVaR}}(\bar{\bm{\omega}}_{Q,x}^*,\bar{\bm{b}}_{Q,x}^*,\bar{\bm{s}}_{Q,x}^*) \\
&\quad= \underbrace{\big[G_{P_x}^{\mathrm{CVaR}}(\bar{\bm{\omega}}_{P,x}^*,\bar{\bm{b}}_{P,x}^*,\bar{\bm{s}}_{P,x}^*)-G_{Q_x}^{\mathrm{CVaR}}(\bar{\bm{\omega}}_{P,x}^*,\bar{\bm{b}}_{P,x}^*,\bar{\bm{s}}_{P,x}^*)\big]}_{\text{Term I}}\\
&\qquad+\underbrace{\big[G_{Q_x}^{\mathrm{CVaR}}(\bar{\bm{\omega}}_{P,x}^*,\bar{\bm{b}}_{P,x}^*,\bar{\bm{s}}_{P,x}^*) - G_{Q_x}^{\mathrm{CVaR}}(\bar{\bm{\omega}}_{Q,x}^*,\bar{\bm{b}}_{Q,x}^*,\bar{\bm{s}}_{Q,x}^*)\big]}_{\text{Term II}}\\
&\qquad+\underbrace{\big[G_{Q_x}^{\mathrm{CVaR}}(\bar{\bm{\omega}}_{Q,x}^*,\bar{\bm{b}}_{Q,x}^*,\bar{\bm{s}}_{Q,x}^*) - G_{P_x}^{\mathrm{CVaR}}(\bar{\bm{\omega}}_{Q,x}^*,\bar{\bm{b}}_{Q,x}^*,\bar{\bm{s}}_{Q,x}^*)\big]}_{\text{Term III}}.
\end{align*}

For Term~I, applying~\eqref{eq:cvar_pointwise} at the feasible point $(\bar{\bm{\omega}}_{P,x}^*,\bar{\bm{b}}_{P,x}^*,\bar{\bm{s}}_{P,x}^*)$ yields
\[
\text{Term I} \le \left(1+\frac{\Gamma}{2\sqrt{1-\beta}}\right)\mathcal{W}_2(P_x,Q_x).
\]

For Term~II, by the optimality of $(\bar{\bm{\omega}}_{Q,x}^*,\bar{\bm{b}}_{Q,x}^*,\bar{\bm{s}}_{Q,x}^*)$ for $G_{Q_x}^{\mathrm{CVaR}}$ over the feasible set, we have
\[
\text{Term II} = G_{Q_x}^{\mathrm{CVaR}}(\bar{\bm{\omega}}_{P,x}^*,\bar{\bm{b}}_{P,x}^*,\bar{\bm{s}}_{P,x}^*) - G_{Q_x}^{\mathrm{CVaR}}(\bar{\bm{\omega}}_{Q,x}^*,\bar{\bm{b}}_{Q,x}^*,\bar{\bm{s}}_{Q,x}^*) \le 0.
\]

For Term~III, applying~\eqref{eq:cvar_pointwise} at the feasible point $(\bar{\bm{\omega}}_{Q,x}^*,\bar{\bm{b}}_{Q,x}^*,\bar{\bm{s}}_{Q,x}^*)$ yields
\[
\text{Term III} \le \left(1+\frac{\Gamma}{2\sqrt{1-\beta}}\right)\mathcal{W}_2(P_x,Q_x).
\]

Combining the three terms, we obtain
\[
G_{P_x}^{\mathrm{CVaR}}(\bar{\bm{\omega}}_{P,x}^*,\bar{\bm{b}}_{P,x}^*,\bar{\bm{s}}_{P,x}^*) - G_{P_x}^{\mathrm{CVaR}}(\bar{\bm{\omega}}_{Q,x}^*,\bar{\bm{b}}_{Q,x}^*,\bar{\bm{s}}_{Q,x}^*) \le 2\left(1+\frac{\Gamma}{2\sqrt{1-\beta}}\right)\mathcal{W}_2(P_x,Q_x).
\]
Taking the expectation on both sides with respect to $X$ yields
\begin{align*}
\mathbb{E}_X \left[ \big| G_{P_X}^{\mathrm{CVaR}}(\bar{\bm{\omega}}_{P,X}^*,\bar{\bm{b}}_{P,X}^*,\bar{\bm{s}}_{P,X}^*) - G_{P_X}^{\mathrm{CVaR}}(\bar{\bm{\omega}}_{Q,X}^*,\bar{\bm{b}}_{Q,X}^*,\bar{\bm{s}}_{Q,X}^*)\big| \right]\le 2\left(1+\frac{\Gamma}{2\sqrt{1-\beta}}\right) \mathbb{E}_X \left[ \mathcal{W}_2(P_X,Q_X)\right]. 
\end{align*}
This completes the proof.
\endproof


\subsection{Proof of Theorem \ref{thm:W2_error_bound}}\label{sec:proof_W2_bound}
Our $\WassD$ error bound in Theorem \ref{thm:W2_error_bound} is built on the following KL-divergence error bound between the true and learned return distributions, whose proof is postponed after the proof of Theorem \ref{thm:W2_error_bound}.

\begin{lemma}\label{lemma:KL_error_bound}
    Under Assumption \ref{assump:diffusion_model}, the following KL-divergence error bound holds:
    \begin{align*}
        &\mE_{x\sim X_t}\left[\KL\big(\tarDist || \genDist\big)\right]\le  C_{KL}\bigg(e^{-\frac{N}{2}(\eta_1 + \eta_N)} \cdot \big(M_3^{2/3} + D\big) + \frac{1}{N}\cdot \big(M_3^{2/3} + D + DL_u\big) + \frac{\varepsilon_{\text{noise}}^2}{1 - e^{- \eta_1}}\bigg),
    \end{align*}
    for some positive constant $C_{\KL}$ independent of $N, D, M_3, L_u$, and $\varepsilon_{\text{noise}}$.
\end{lemma}

We next prove Theorem \ref{thm:W2_error_bound}. 
The general idea is to convert the KL bound in Lemma~\ref{lemma:KL_error_bound} into a TV (total variation) bound via Pinsker's inequality, and subsequently bound the $\WassD$ distance by the TV distance. An alternative approach is to directly convert the KL bound into a $\WassD$ bound, provided that the generated return distribution \(Q_{x}\) satisfies e.g. Talagrand's transportation cost inequality \citep{otto2000generalization, bolley2005weighted}. However, this alternative requires stringent conditions (such as Gaussian tails) on the generated return distribution and
imposes severe structural restrictions on the trained neural network. Consequently, we do not adopt this approach in this work.

\begin{proof}{Proof of Theorem \ref{thm:W2_error_bound}.}
Firstly, for any given $X_t= x$, denote by $\TV\big(\tarDist || \genDist\big)$ the TV distance between $\tarDist$ and $\genDist$. According to Theorem 6.15 of \cite{villani2009optimal}, for any positive constant $C_B$ (which we will optimize later), we have
\begin{align*}
    &\WassD\big(\tarDist || \genDist\big) \\
    &\le \bigg(C^2_B\TV\big(\tarDist || \genDist\big) + \int_{\|r\|>C_B}\|r\|^2\tarDist(\intd r) + \int_{\|r\|>C_B}\|r\|^2\genDist(\intd r)\bigg)^{1/2}\\
    &\le \bigg(C^2_B\TV\big(\tarDist || \genDist\big) + \frac{1}{C_B}\mE_{R_{t + 1}\sim P_x}\big[\|R_{t + 1}\|^3\big] + \frac{1}{C_B}\mE_{\tilde{R}\sim \genDist} \big[\|\tilde{R}\|^3\big]\bigg)^{1/2}\\
    &\le \bigg(C^2_B\TV\big(\tarDist || \genDist\big)\bigg)^{1/2} + \bigg(\frac{1}{C_B}\mE_{R_{t + 1} \sim \tarDist}\big[\|R_{t + 1}\|^3\big]\bigg)^{1/2} + \bigg(\frac{1}{C_B}\mE_{\tilde{R}\sim \genDist}\big[\|\tilde{R}\|^3\big]\bigg)^{1/2},
\end{align*}
where we use the inequality $\mE\big[\|\bU\|^2\cdot\mathbf{1}_{\|\bU\|>C_B}\big] \le \frac{1}{C_B} \mE \big[\|\bU\|^3\big]$ for any random variable $\bU$ and we use the fact that $\sqrt{w_1 + w_2 + w_3} \le \sqrt{w_1} + \sqrt{w_2} + \sqrt{w_3}$ for any $w_1, w_2, w_3 \ge 0$. Taking expectation on both sides w.r.t. $x\sim X_t$ and applying Jensen's inequality, we obtain
\begin{align}
    &\mE_{x\sim X_t}\big[\WassD\big(\tarDist || \genDist\big) \big]\nonumber\\
    &\le C_B\bigg(\mE_{x\sim X_t}\big[\TV\big(\tarDist || \genDist\big)\big]\bigg)^{1/2} + \frac{1}{\sqrt{C_B}}\bigg(\mE_{x\sim X_t}\big[\mE_{R_{t + 1}\sim\tarDist}\big[\|R_{t + 1}\|^3\big]\big]\bigg)^{1/2}\nonumber\\
    &\qquad + \frac{1}{\sqrt{C_B}}\bigg(\mE_{x\sim X_t}\big[\mE_{\tilde{R}\sim\genDist}\big[\|\tilde{R}\|^3\big]\big]\bigg)^{1/2}\nonumber\\
    &\le C_B\bigg(\mE_{x\sim X_t}\big[\TV\big(\tarDist || \genDist\big)\big]\bigg)^{1/2} + \frac{2\sqrt{M_3}}{\sqrt{C_B}}, \label{eq: TV_to_W2}
\end{align}
where the last inequality is due to Assumption \ref{assump:diffusion_model}.

Next, by Pinsker's inequality and Jensen's inequality, 
\begin{align*}
    \mE_{x\sim X_t}\big[\TV\big(\tarDist || \genDist\big)\big] \le \frac{1}{\sqrt{2}}\bigg( \mE_{x\sim X_t}\big[\KL\big(\tarDist || \genDist\big)\big]\bigg)^{1/2}.
\end{align*}
Substituting into \eqref{eq: TV_to_W2} and using Lemma \ref{lemma:KL_error_bound}, we have
\begin{align}\label{eq:rhs}
    \mE_{x\sim X_t}\big[\WassD\big(\tarDist || \genDist\big) \big]\le C_B\cdot (C_{\KL}/2)^{1/4}\cdot \mathcal{E}^{1/4} + \frac{2\sqrt{M_3}}{\sqrt{C_B}}.
\end{align}
where for notation simplicity we define 
\begin{align*}
    \mathcal{E} := e^{-\frac{N}{2}(\eta_1 + \eta_N)} \cdot \big(M_3^{2/3} + D\big) + \frac{1}{N}\cdot \big(M_3^{2/3} + D + DL_u\big) + \frac{\varepsilon_{\text{noise}}^2}{1 - e^{- \eta_1}}.
\end{align*}

To minimize the error bound on the right-hand side of \eqref{eq:rhs}, we choose
\begin{align*}
    C_B = (C_{\KL}/2)^{-1/6}\cdot M_3^{1/3}\cdot\mathcal{E}^{-1/6}.
\end{align*}
which yields
\begin{align*}
    \mE_{x\sim X_t}\big[\WassD\big(\tarDist || \genDist\big) \big]\le C_{\WassD} \cdot M_3^{1/3}\cdot \mathcal{E}^{1/12},
\end{align*}
where $C_{\WassD} := 3\cdot (C_{\KL}/2)^{1/12} > 0$ is independent of $N, D, M_3, L_u$ and $\varepsilon_{\text{noise}}$. This completes the proof.







\end{proof}

\begin{proof}{Proof of Lemma \ref{lemma:KL_error_bound}.}

To prove Lemma \ref{lemma:KL_error_bound}, the idea is to treat DDPM as a discretized continuous-time SDE (stochastic differential equation), and cast the discrete process into the framework of score-based continuous-time conditional diffusion models (SCDMs) \citep{songscore}. Under this formulation, the proof proceeds by applying the KL error bound for SCDMs from Proposition 2 of \cite{gao2026data}.



Conditional on $X_t=x$, SCDMs consider the forward SDE: 
\begin{align*}
    \intd\bY(\tau) = -f_{DM}(\tau)\bY(\tau)\intd\tau + g_{DM}(\tau) \intd\bB(\tau),\ \bY(0)\sim \tarDist,\ \tau\in [0, T_g],
\end{align*}
where $\bB$ is a standard Brownian motion in $\mR^{D}$ independent of $\bY(0)$, $f_{DM}(\tau): = \frac{1}{2}\beta_{DM}(\tau),\ g_{DM}(\tau): = \sqrt{\beta_{DM}(\tau)},\ \beta_{DM}(\tau) := a\tau + b$ for some constants $a, b$. Given $X_t=x$, denote by $q_\tau(\cdot | x)$ marginal density of $\bY(\tau)$ at time $\tau$, which is unknown due to the unknown $\bY(0)\sim\tarDist$. However, conditional on $\bY(0) = r_{t+1}$, $\bY(\tau)$ is Gaussian and known, whose density we denote by $q_{\tau|0}(\cdot |r_{t+1}, x)$. To generate new samples, SCDMs consider the time-reversed process $\{\tilde\bY(\tau):= \bY(T_g - \tau), \tau\in[0, T_g]\}$, which satisfies a reverse SDE starting from $\tilde\bY(0)\sim q_{T_g}(\cdot | x)$:
\begin{align*}
    \intd\tilde{\bY}(\tau) = [f_{DM}(T_g - \tau)\tilde\bY(\tau) + g^2_{DM}(T_g - \tau) \nabla\log q_{T_g - \tau}(\tilde \bY(\tau) | x)]\intd\tau + g_{DM}(T_g - \tau) \intd\tilde\bB(\tau),
\end{align*}
where $\nabla\log q_{\tau}(\cdot | x)$ is the unknown (conditional) score function and needs to be learned. 

With the noise scheduler $\{\eta_n\}_{n=1}^N$ from Section \ref{sec:ddpm}, setting
\begin{align}\label{eq:DDPM_setting}
    T_g = 1,\quad a = N\cdot(\eta_N - \eta_1),\quad b = \eta_1 \cdot N, \quad \Delta\tau :=\frac{1}{N},
\end{align}
the DDPM forward process \eqref{eq:forward} (resp. reverse process \eqref{eq:reverse}) can be viewed as a discretization of the forward (resp. reverse) SDE; see Appendix B of \cite{songscore} for details. For the unknown score function, SCDMs aim to train a score network $s_\theta$ that minimizes the score-matching (SM) loss
\begin{align*}
    L_{SM}(\theta) := \frac{1}{N}\sum_{n=1}^N\mathbb{E}_{x\sim X_{t}}\left[\mE_{R_{t+1}^{(n)}\sim q_{\frac{n}{N}}(\cdot | x)}\left\| \nabla\log q_{\frac{n}{N}}(R^{(n)}_{t+1} | x) - s_\theta \left(R^{(n)}_{t+1},n;x\right)\right\|^2 \right].
\end{align*}
Since $L_{SM}(\theta)$ is intractable, a standard surrogate is to minimize the denoising score-matching (DSM) loss
\begin{align*}
    L_{DSM}(\theta) = \frac{1}{N}\sum_{n=1}^N\mathbb{E}_{x\sim X_{t}}\left[\mE_{(R_{t+1}^{(0)}, R_{t+1}^{(n)})}\left\| \nabla\log q_{\frac{n}{N}|0}(R^{(n)}_{t+1} | R_{t+1}^{(0)}, x) - s_\theta \left(R^{(n)}_{t+1},n;x\right)\right\|^2 \right],
\end{align*}
which differs from $L_{SM}(\theta)$ only by a $\theta$-independent constant \citep{vincent2011connection, zhang2025heuristic}. The DSM loss is in turn a reweighted version of our DDPM noise prediction loss $L(\theta)$ in \eqref{eq:loss}: according to (12)-(14) of \cite{yang2023diffusion},
\begin{align}
    &\mathbb{E}_{(X_{t},R_{t+1}),\epsilon}\left[\left\|\bigg( \frac{-\epsilon}{\alpha_{\frac{n}{N}}}\bigg) - \frac{-\bepsilon_\theta \left(R^{(n)}_{t+1},n;X_{t}\right)}{\alpha_{\frac{n}{N}}}\right\|^2 \right]\nonumber\\
    &=\mathbb{E}_{(X_{t},R_{t+1}), \epsilon}\left[\left\| \nabla\log q_{\frac{n}{N}|0}(R^{(n)}_{t+1} | R_{t+1}, X_t) - s_\theta \left(R^{(n)}_{t+1},n;X_{t}\right)\right\|^2 \right], \label{eq: eps_to_score}
\end{align}
where $\alpha^2_{\tau} := 1 - e^{-\frac{a}{2}\tau^2 - b\tau}$ and the score network is identified as $s_\theta(\cdot, n; x):= -\bepsilon_\theta(\cdot, n; x)/\alpha_{\frac{n}{N}}$. 


We now verify the three conditions required by Proposition 2 of \cite{gao2026data}. 

Firstly, the Lipschitz continuity condition holds by virtue of Assumption \ref{assump:diffusion_model}(1). In fact, conditional on $X_t = x$, the factor model \eqref{eq:model} indicates that $P_x$ admits a twice continuously differentiable positive density, which we denote by $p_{R_{t + 1}}(\cdot | x)$. In addition, 
\begin{align*}
    p_{R_{t + 1}}(r | x) = p_u(r - f(x)), \quad \nabla_r\log p_{R_{t + 1}}(r | x) = \nabla_u \log p_u(u)\big|_{u=r - f(x)},
\end{align*}
which implies that $p_{R_{t + 1}}(\cdot | x)$ is also positive on $\mR^D$ and $\nabla_r\log p_{R_{t + 1}}(\cdot | x)$ is $L_u$-Lipschitz continuous with constant $L_u$ free of $x$. 

Next, the finite second moment condition follows from Assumption \ref{assump:diffusion_model}(2), since, by Lyapunov's inequality and Jensen's inequalities,
\begin{align*}
    \mE_{x\sim X_t}\big[\mE_{R_{t + 1}|X_t=x}||R_{t + 1}||^2\big] &\le \mE_{x\sim X_t}\big[\big(\mE_{R_{t + 1}|X_t=x}||R_{t + 1}||^3\big)^{2/3}\big]\\
    &\le \big(\mE_{x\sim X_t}\big[\mE_{R_{t + 1}|X_t=x}||R_{t + 1}||^3\big]\big)^{2/3} \\
    &\le M_3^{2/3}.
\end{align*}

Finally, we can obtain an error bound on the SM loss $L_{SM}(\theta)$ from Assumption \ref{assump:diffusion_model}(3). In fact, by \cite[Theorem 4.6]{zhang2025heuristic}, there exists a constant $C_{SM}\ge 0$ that is free of $\theta$ such that for any $\theta$,
\begin{align}\label{eq:SM_to_DSM_loss}
    L_{SM}(\theta) + C_{SM} = L_{DSM}(\theta) \le \alpha^{-2}_{\frac{1}{N}}L(\theta) = \big(1-e^{-\eta_1 - \frac{\eta_N - \eta_1}{N}}\big)^{-1} L(\theta) \le \big(1 - e^{- \eta_1}\big)^{-1}L(\theta), 
\end{align}
where the first inequality is from \eqref{eq: eps_to_score}. Thus, by \eqref{eq:assum_DSM_error} of Assumption \ref{assump:diffusion_model}(3), for our neural network $\epsilon_{\hat\theta^*}$, we have $L_{SM}(\hat\theta^*) \le \big(1 - e^{- \eta_1}\big)^{-1}\varepsilon_{\text{noise}}^2$.

Therefore, applying Proposition 2 of \cite{gao2026data} with settings in \eqref{eq:DDPM_setting}, for some positive constant $C_{SCDM} > 0$ independent of $N, D, L_u, M_3$ and $\varepsilon_{\text{noise}}$, we have 
\begin{align*}
    &\mE_{x\sim X_t}\left[\KL\big(\tarDist || \genDist\big)\right]\nonumber\\
    &\le C_{SCDM} \cdot \bigg(
    e^{-\frac{N}{2}(\eta_1 + \eta_N)}\big(M_3^{2/3} + De^{-\frac{N}{2}(\eta_1 + \eta_N)}\big) + (1 + 1)^3 \cdot \frac{\varepsilon_{\text{noise}}^2}{1 - e^{- \eta_1}}  + \frac{1}{N}(M_3^{2/3} + D)(1 + 1)^5\nonumber\\
    &\qquad + \frac{M_3^{2/3}}{N} (1 + 1)^2 + \frac{D}{N} \big[L_u + (1 + 1)^4 \big]+ \frac{D}{N^2}(L_u + 1 + 1)\bigg)\nonumber\\
    &\le C_{KL}\bigg(e^{-\frac{N}{2}(\eta_1 + \eta_N)} \cdot \big(M_3^{2/3} + D\big) + \frac{1}{N}\cdot \big(M_3^{2/3} + D + DL_u\big) + \frac{\varepsilon_{\text{noise}}^2}{1 - e^{- \eta_1}}\bigg)
\end{align*}
where we collect all remaining lower-order terms into a constant $C_{KL} > 0$ that is independent of $N$, $D$, $L_u, M_3$, and $\varepsilon_{\text{noise}}$. This completes the proof.
\end{proof}

\subsubsection{Finite third absolute moment of the generated distribution}\label{sec:gen_dist_finite_moment}

In the following lemma, we provide a sufficient condition under which the learned return distribution $Q_X$ inherits a finite third absolute moment, thereby validating \eqref{eq:assum_gen_moment_3} in Assumption \ref{assump:diffusion_model} used in Theorem \ref{thm:W2_error_bound}. 
Recall the (conditional) DDPMs introduced in Section \ref{sec:ddpm} and the learned return $\widetilde{R}^{(0)}_{t + 1}$ generated by \eqref{eq:reverse}.
Denote by $\mathcal{X}$ the support of the factor matrix $X_t$.

\begin{lemma}\label{lemma:finite_third_moment}
     Suppose that there exist measurable functions $H_0, H_1:\mathcal{X}\to[0,\infty)$ such that for any given condition $X_t=x\in\mathcal{X}$, the trained neural network $\bepsilon_\theta(\cdot, n; x)$ satisfies the linear growth condition
    \begin{align}\label{eq: eps_linear_growth}
        \big\|\epsilon_{\theta}(r,n;x)\big\| \le H_1(x) \|r\|+H_0(x)\qquad\forall r\in\mathbb{R}^{D}, n=1,\cdots, N.
    \end{align}
    Assume in addition that $H_0, H_1$ satisfy that
    \begin{align}\label{eq:growth_integ_condition}
        \mathbb{E}_{x\sim X_t}\big[e^{3C_{K}H_1(x)}[ H_0(x)^{3} + 1]\big]<\infty,
    \end{align}
    where $C_{K}:=\sum_{n=1}^{N}\eta_{n}(1-\bar\zeta_{n})^{-1/2}$. Then, there exists a positive constant $C_\eta$ that is only dependent on the noise scheduler $\{\eta_n\}_{n=1}^N$ such that 
    \begin{align*}
        \mE_{x\sim X_t} \big[\mathbb{E}_{\widetilde{R}^{(0)}_{t+1} \sim Q_x}\|\widetilde R^{(0)}_{t+1}\|^{3}\big] \le C_\eta\mE_{x\sim X_t}\bigg[e^{3C_{K} H_1(x)}\big[(H_0(x))^3 + M_G + 1\big]\bigg],
    \end{align*}
    where we denote the third absolute moment of standard Gaussian variable by $M_G := \mE_{\epsilon\sim\mathcal{N}(0, I_D)}[\|\epsilon\|^3] = 2\sqrt{2}\cdot \frac{\Gamma((D + 3)/2)}{\Gamma(D/2)}$ with Gamma function $\Gamma(w) := \int_{0}^{\infty}u^{w - 1}e^{-u}\intd u$.
\end{lemma}

The linear growth condition \eqref{eq: eps_linear_growth} imposed on the trained neural network is relatively mild and standard in the literature. Indeed, the global Lipschitz continuity of neural networks—which directly implies linear growth—has been studied extensively. This property holds generally whenever the network architecture comprises Lipschitz-continuous activation functions (e.g., ReLU, SiLU, tanh, and sigmoid) and weight matrices with bounded operator norms \citep{szegedy2013intriguing}. Furthermore, the integrability condition \eqref{eq:growth_integ_condition} governs the probabilistic behavior of the state-dependent coefficients \(H_1(X_t)\) and \(H_0(X_t)\). This condition simplifies significantly under standard settings; it is automatically satisfied if \(H_0\) and \(H_1\) are constants independent of the conditioning variable \(X_{t}\). In a more general case where \(H_1(x)\) remains a constant and \(H_0(x)\) exhibits at most linear growth in \(x\), condition \eqref{eq:growth_integ_condition} reduces to requiring the finiteness of the third absolute moment of \(X_{t}\), which is a standard and non-restrictive assumption. We next prove Lemma \ref{lemma:finite_third_moment}.

\begin{proof}{Proof of Lemma \ref{lemma:finite_third_moment}.}
We first rewrite the reverse process \eqref{eq:reverse} as
\begin{align}\label{eq: reverse_recursion}
    \widetilde R^{(n-1)}_{t+1} = A_{n}\widetilde R^{(n)}_{t+1} + B_{n} \epsilon_{\theta}\big(\widetilde R^{(n)}_{t+1},n;X_{t}\big) + \sigma_{n}  \epsilon_{n},\quad \epsilon_{n}\sim\mathcal{N}(0,I_{D}), 
\end{align}
where $A_{n} := \frac{1}{\sqrt{\zeta_{n}}},B_{n} := - \frac{\eta_{n}}{\sqrt{\zeta_{n}(1-\bar\zeta_{n}})}, \sigma_{n}^{2}=\frac{1-\bar\zeta_{n-1}}{1-\bar\zeta_{n}} \eta_{n}$. Throughout the proof, we consider the conditional third absolute moment function
\begin{align*}
    m_{n}(x):=\mathbb{E}_{\widetilde{R}^{(n)}_{t+1}|X_t=x}\big[\|\widetilde R^{(n)}_{t+1}\|^{3} \big],\quad n=0, 1,\cdots, N.
\end{align*}
Applying Minkowski's inequality to \eqref{eq: reverse_recursion} and invoking the linear-growth assumption \eqref{eq: eps_linear_growth}, we obtain
\begin{align}
    \big(m_{n - 1}(x)\big)^{1/3} &\le A_n  \big(m_{n}(x)\big)^{1/3} + |B_n|\bigg(\mE_{\widetilde{R}^{(n)}_{t+1}|X_t=x}\big[\big\|\epsilon_{\theta}\big(\widetilde R^{(n)}_{t+1},n; x\big)\big\|^3\big]\bigg)^{1/3} + \sigma_n M_G^{1/3}\nonumber\\
    &\le A_n  \big(m_{n}(x)\big)^{1/3} + |B_n|\bigg(\mE_{\widetilde{R}^{(n)}_{t+1}|X_t=x}\big[\big(H_1(x)\big\|\widetilde{R}^{(n)}_{t+1}\big\| + H_0(x)\big)^3\big]\bigg)^{1/3} + \sigma_n M_G^{1/3}\nonumber\\
    &\le A_n  \big(m_{n}(x)\big)^{1/3} + |B_n|\big[H_1(x)\big(m_{n}(x)\big)^{1/3} + H_0(x)\big] + \sigma_n M_G^{1/3}\nonumber\\
    &\le \kappa_n(x)\big(m_{n}(x)\big)^{1/3} + \nu_n(x). \label{eq: norm_iteration}
\end{align}
where we recall $M_G = \mE_{\epsilon\sim\mathcal{N}(0, I_D)}[\|\epsilon\|^3] = \mE[\|\epsilon_n\|^3]$ and we define
\begin{align*}
    \kappa_{n}(x) := A_{n}+|B_{n}| H_1(x) ,\quad \nu_{n}(x) := |B_{n}| H_0(x)+\sigma_{n} M_G^{1/3}.
\end{align*}

Iterating the recursion \eqref{eq: norm_iteration} through the $N$ generation steps, we obtain
\begin{align*}
    \big(m_{0}(x)\big)^{1/3} \le \Pi_{N}(x) \big(m_{N}(x)\big)^{1/3}+\sum_{n=1}^{N}\nu_{n}(x)\Pi_{n - 1}(x),
\end{align*}
where $\Pi_{n}(x) := \prod_{k=1}^{n}\kappa_{k}(x)$ for $n=1, \cdots, N$ and $\Pi_0(x) = 1$. Applying the fact $(w_1 + w_2)^3 \le 4(w_1^3 + w_2^3)$ for $w_1, w_2 \ge 0$ produces
\begin{align}\label{eq: moment_bound_ineq}
    \mE_{x\sim X_t}[m_{0}(x)] \le 4\mE_{x\sim X_t}\big[(\Pi_{N}(x))^3m_{N}(x)\big] + 4\mE_{x\sim X_t}\Bigg[\bigg(\sum_{n=1}^{N}\nu_{n}(x)\Pi_{n - 1}(x)\bigg)^3\Bigg].
\end{align}

We now bound the two expectations on the right-hand side of \eqref{eq: moment_bound_ineq} separately.

For the first term, note that for any $X_t=x$, we have $\big(\widetilde{R}^{(N)}_{t+1}|X_t=x\big)\sim\mathcal{N}(0, I_D)$, so
\begin{align*}
    m_{N}(x)= \mathbb{E}_{\widetilde{R}^{(N)}_{t+1}|X_t=x}\big[\|\widetilde R^{(N)}_{t+1}\|^{3}\big] = M_G.
\end{align*}
On the other hand, using $\prod_{n=1}^{N}\zeta_{n}^{-1/2}=\bar\zeta_{N}^{-1/2}$ and $\log(1+w)\le w$ for any $w \ge -1$, we have 
\begin{align}\label{eq: prod_bound}
    \Pi_{N}(x)  = \prod_{n=1}^N\frac{1}{\sqrt{\zeta_n}}\bigg(1 + \frac{\eta_nH_1(x)}{\sqrt{1 - \bar\zeta_n}}\bigg) \le \frac{1}{\sqrt{\bar\zeta_{N}}} \exp\big(C_{K} H_1(x)\big).
\end{align}
where we recall that $C_{K} = \sum_{n=1}^{N}\frac{\eta_{n}}{\sqrt{1-\bar\zeta_{n}}}$. Consequently,
\begin{align}\label{eq: m_bound_1}
    \mE_{x\sim X_t}\big[(\Pi_{N}(x))^3m_{N}(x)\big]  \le \frac{M_G}{\bar \zeta_N^{3/2}}\mE_{x\sim X_t}\big[e^{3C_{K} H_1(x)}\big].
\end{align}

We next bound the second term in \eqref{eq: moment_bound_ineq}. We first recall the weighted power-mean inequality: for non-negative constants $c_{1},\dots,c_{N}$ and strictly positive weights $w_{1},\dots,w_{N}$ with $W:=\sum_{n}w_{n}$, 
\begin{align*}
    \Big(\sum_{n=1}^{N}c_{n}\Big)^{3}=W^{3}\Big(\sum_{n=1}^{N}\frac{w_{n}}{W}\cdot\frac{c_{n}}{w_{n}}\Big)^{3}\le W^{3}\sum_{n=1}^{N}\frac{w_{n}}{W}\cdot\frac{c_{n}^{3}}{w_{n}^{3}}=W^{2}\sum_{n=1}^{N}\frac{c_{n}^{3}}{w_{n}^{2}}.
\end{align*}

We set $c_{n}=\nu_{n}(x) \Pi_{n-1}(x)$ and weights $w_{n}=(\Pi_{n-1}(x))^{3/2}$, so that $c_{n}^{3}/w_{n}^{2} = (\nu_{n}(x))^{3}$. Then,
\begin{align}\label{eq: m_bound_2}
    \Bigg(\sum_{n=1}^{N}\nu_{n}(x) \Pi_{n-1}(x)\Bigg)^{3} \le \Bigg(\sum_{n=1}^{N}(\Pi_{n-1}(x))^{3/2}\Bigg)^{2} \cdot \sum_{n=1}^{N}(\nu_{n}(x))^{3}.
\end{align}

Since $\Pi_{n-1}(x) =\Pi_{N}(x)\big/\prod_{k=n}^{N}\kappa_{k}(x)$ and $\kappa_{k}(x) \ge A_{\min} :=\min_{n}A_{n}=(1- \eta_{1})^{-1/2} > 1$, we have 
\begin{align}\label{eq: m_bound_2_1}
    \sum_{n=1}^{N}(\Pi_{n-1}(x))^{3/2} \le \sum_{n=1}^{N}(\Pi_{N}(x))^{3/2} A_{\min}^{-\frac{3}{2}(N-n+1)} \le \frac{(\Pi_{N}(x))^{3/2}}{A_{\min}^{3/2} - 1}.
\end{align}

On the other hand, from $\nu_n(x) = |B_{n}| H_0(x)+\sigma_{n} M_G^{1/3}$, we know that
\begin{align}\label{eq: m_bound_2_2}
    \sum_{n=1}^{N}(\nu_{n}(x))^{3} \le 4\big[(H_0(x))^3 + M_G\big]\sum_{n=1}^N(|B_n|^3 + \sigma_n^3):= 4C_\nu\big[(H_0(x))^3 + M_G\big].
\end{align}

Substituting \eqref{eq: m_bound_2_1} and \eqref{eq: m_bound_2_2} into \eqref{eq: m_bound_2} and using \eqref{eq: prod_bound} and taking expectation w.r.t. $x\sim X_t$, we have
\begin{align}
    \mE_{x\sim X_t}\bigg[\Big(\sum_{n=1}^{N}\nu_{n}(x) \Pi_{n-1}(x)\Big)^{3}\bigg] &\le \mE_{x\sim X_t}\bigg[\frac{e^{3C_{K} H_1(x)}}{\bar\zeta_N^{3/2}(A_{\min}^{3/2} - 1)^2}\cdot 4C_\nu\big[(H_0(x))^3 + M_G\big]\bigg]\nonumber\\
    &= \frac{4C_\nu}{\bar\zeta_N^{3/2}(A_{\min}^{3/2} - 1)^2}\mE_{x\sim X_t}\bigg[e^{3C_{K} H_1(x)}\big[(H_0(x))^3 + M_G\big]\bigg].\label{eq: m_bound_2_final}
\end{align}

Finally, combining \eqref{eq: moment_bound_ineq}, \eqref{eq: m_bound_1} and \eqref{eq: m_bound_2_final}, we conclude that
\begin{align*}
    &\mE_{x\sim X_t}[m_{0}(x)]  = \mE_{x\sim X_t}\big[\mathbb{E}_{\tilde{R} \sim \genDist} \big[\|\tilde{R} \|^{3}  \big]\big] \\
    &\le \frac{4M_G}{\bar \zeta_N^{3/2}}\mE_{x\sim X_t}\big[e^{3C_{K} H_1(x)}\big] + \frac{16C_\nu}{\bar\zeta_N^{3/2}(A_{\min}^{3/2} - 1)^2}\mE_{x\sim X_t}\bigg[e^{3C_{K} H_1(x)}\big[(H_0(x))^3 + M_G\big]\bigg]\\
    &\le C_\eta\mE_{x\sim X_t}\bigg[e^{3C_{K} H_1(x)}\big[(H_0(x))^3 + M_G + 1\big]\bigg],
\end{align*}
with $C_\eta = 4\bar\zeta_N^{-3/2}\big[1 + 4C_\nu((1-\eta_1)^{-3/4} - 1)^{-2}\big]$ only dependent on the noise scheduler $\{\eta_n\}_{n=1}^N$. This completes the proof. 
\end{proof}

\end{document}